\documentstyle[twocolumn]{article}
\begin{document}
\begin{titlepage}
\title{Low Field Magnetic Response of the Granular Superconductor La$_{1.8}$Sr$_{0.2}$CuO$_4$}
\author{L. Leylekian, M. Ocio and L. A. Gurevich \thanks{Permanent address: Institute of Solid State Physics, 
Chernogolovka, Moscow Region, 142432, Russia. Current address: Delft 
University of Technology, Lorentzweg 1, 2628 CJ Delft, The Netherlands.}
\\Service de Physique de 
l'Etat Condens\'e, CEA,\\ CE Saclay, 91191 Gif sur Yvette Cedex
\and M. V. Feigel'man
\\Landau Institute for Theoretical Physics,
Moscow, 117940, Russia.}

\maketitle
\begin{abstract}
The properties of the low excitation field magnetic response of the granular 
high temperature (HT$_c$) superconductor La$_{1.8}$Sr$_{0.2}$CuO$_4$~ have 
been analyzed at low temperatures. The 
response of the Josephson currents has been extracted from the data. It is 
shown that intergrain current response is fully irreversible, producing 
shielding response, but do not 
carry Meissner magnetization. Analysis of 
the data shows that the system of Josephson currents freezes into a glassy 
state
even in the absense of external magnetic field, which is argued to be a 
consequence of the $d$-wave nature of superconductivity in La$_{1.8}$Sr$_{0.2}$CuO$_4$.
The macroscopic diamagnetic response to very weak variations of the magnetic
field is shown to be strongly irreversible but still qualitatively different 
 from 
 any previously known kind of the critical-state behaviour in superconductors. 
 A phenomenological
description of these data is given in terms of a newly proposed
 ``fractal'' model of irreversibility in superconductors.

PACS numbers 74.50.+r, 74.60.Ge, 74.80.Bj
\end{abstract}
\end{titlepage}

\input epsf

\setcounter{dbltopnumber}{3}
\setcounter{section}{0}
\setcounter{subsection}{0}
\setcounter{subsubsection}{0}
\setcounter{equation}{0}

\section{Introduction}
Granular superconductors (SC) are composed of a very large number of
small (micron-size) superconductive grains which are coupled together
due to the Josephson  tunnelling (or, in some cases, due to
the proximity effect). These systems are inherently disordered due to
randomness in the sizes of grains and in their mutual distances. Usually
the strength of Josephson coupling between grains is rather weak, so
the maximum Josephson energy of the contact between two grains is much below
the intragrain superconductive condensation energy.
Therefore granular SC can be considered as systems with a
two-level organization: their short-scale properties are determined by the
superconductivity of individual grains, whereas the macroscopic SC behaviour
is governed by the weak intergrain couplings. In the treatment of the latter,
one can neglect any internal structure of SC grains and describe them just
by the phases $\phi_j$ of their superconductive order parameters $\Delta_j =
|\Delta|_j \exp(i\phi_j)$. As a result, the macroscopic behaviour of granular
SC can be described by a classical free energy functional of the form
(cf. Ref.\ \cite{Blatter,Lubens,Vinok}):
\begin{eqnarray}
H = \frac{1}{2} \sum_{ij} E_J^{ij} cos(\phi_i- \phi_j - \alpha_{ij})+\nonumber \\
 \int d^3r \left(\frac{1}{8\pi} (rot{\bf A})^2 - \frac{1}{4\pi}
rot{\bf A}\cdot{\bf H}_{ext}\right)
\label{ham}
\end{eqnarray}
where $\alpha_{ij}= \frac{2\pi}{\Phi_0} \int_i^j {\bf A} d{\bf r}$ is the phase
difference induced by the electromagnetic vector potential
$\bf A$ and $\Phi_0=\pi\hbar c/e\,$,
whereas the  coupling strengths $E_J^{ij}$ are proportional to the maximum
Josephson currents: $E_J^{ij} = \frac{\hbar}{2e} I^c_{ij}\,$. The vector potential
$\bf A$ in Eq.(\ref{ham}) is the sum of the vector potential ${\bf A}_{ext}$
of the external magnetic field ${\bf H}_{ext}$ and of the Josephson
currents-induced vector potential ${\bf A}_{ind}$. In the absence of external
magnetic field, the lowest-energy state for the ``Hamiltonian'' (\ref{ham})
is, clearly, a macroscopically superconductive state with all phases $\phi_j$
equal to each other. Thus that granular SC system looks similar to the random
XY ferromagnet with randomness in the values of the coupling strengths
$E_J^{ij}$'s  (apart from the possible role of the induced vector potential
${\bf A}_{ind}$ which will be dicussed later);
within this analogy  the role of XY ``spin components'' is taken by
${S_x= \cos\phi_j}$, ${S_y= \sin\phi_j}$.

The situation becomes a lot more complicated
in the presence of non-zero external magnetic field ${\bf H}_{ext}$, which makes the
system {\it randomly frustrated} (since magnetic fluxes penetrating plaquettes
between neighbouring grains are random fractional parts of $\Phi_0$). When
the external field is sufficiently strong, $H_{ext} \gg H_0 = \Phi_0/a_0^2$
(here $a_0$ is the characteristic intergrain distance), the random
phases $\alpha_{ij}$ become of the order of $\pi$ or larger, which means
complete frustration of the intergrain couplings -- i.e. the system is
then expected to resemble the XY {\it spin glass}.
Actually the random Josephson network in a
magnetic field is not exactly identical to the XY spin glass due to
the following reasons \cite{Blatter}:
i). The effective couplings ${\tilde E}_J^{ij} = E_J^{ij} \exp(i\alpha_{ij})$
between ``spins'' ${\bf S_i}$ of the frustrated SC network are
{\it random complex} numbers
whereas in the XY spin glass model, they are {\it real random} numbers.
ii). Generally
the phases $\alpha_{ij}$ depend  on the total magnetic induction ${\bf B} =
{\bf H}_{ext} + {\bf B}_{ind} $, i.e. the effective couplings $\tilde{E}_J^{ij}$
 depend on the phase variables $\phi_j$ determining the intergrain currents
$ I_{ij} = I_{ij}^c sin(\phi_i-\phi_j - \alpha_{ij})$.
In some cases the effects produced by the self-induced magnetic field
${\bf B}_{ind}$ are weak and can be neglected (the quantitative criterion
will be discussed later on), so that phases $\alpha_{ij}$ can be considered
as being fixed by the external field.

The model described by the
Hamiltonian (\ref{ham}) with fixed $\alpha_{ij}$'s and $H_{ext} \gg H_0$
is usually called ``gauge glass'' model.  It is expected on the basis
of the analytical \cite{Lubens,Vinok,fi,Feigel} as well
as numerical \cite{young,huse} results that the gauge glass model in 3D space
exhibits a true phase transition into a low-temperature glassy
superconductive (nonergodic) state. The mean-field theory of such a
low-temperature state shows \cite{Vinok,Feigel} that it is characterized
by the presense of a finite effective
penetration depth for the {\it variation} of an external field, nonzero
macroscopic critical current, and the absense of a macroscopic Meissner
 effect.
The full model (\ref{ham}) with $\alpha$'s containing contribution from
${\bf B}_{ind}$ is sometimes called ``gauge glass with screening''
\cite{screened}.
The effect of screening on the presence and properties of the phase transition
into a glassy state is not completely clear;  some numerical results
\cite{screened}
indicate the absence of a true phase transtion in a 3D model with screening.
Quantitatively, the strengh of screening is determined by the ratio
$\beta_L = 2\pi{\cal L}I_c/c\Phi_0$ where ${\cal L}$ is the characteristic
inductance of an elementary intergrain current loop \cite{Vela}. In the ceramics
with $\beta_L \ll 1\,$, screening effects become important on a long-distance
scale $\sim a_0/\sqrt{\beta_L}$ only (i.e. they are similar to the strongly
type-II superconductors with disorder).

  Apart from its relevance for the description of granular superconductors,
the gauge glass model with screening is rather often considered
(e.g. Ref.\ \cite{Fisher}) as a simplified model describing the large-scale
behaviour of disordered bulk type-II superconductors
in the mixed state (so-called vortex glass problem).
Actually it is unclear a priori how these two
problems are related; an obvious difference between them is that the basic
ingredient of the latter is the vortex lattice which is clearly
an anisotropic object, whereas the former does not contain any prescribed
direction in the 3D space. On the other hand, the granular superconductor in a
moderate magnetic field $H_{ext} \leq H_0$ may be considered as a kind of
disordered type-II superconductor, where the notion of a hypervortex
(which is the macroscopic analogue of the Abrikosov vortex) can be introduced
\cite{Lubens,sonin}. Therefore, the macroscopic properties of a granular network
at $H_{ext} \le H_0$ may
resemble those of the vortex glass; in such a scenario a phase transtion
between vortex glass and gauge glass phases would be expected in a granular
superconductive network at $H_{ext} \sim H_0$ (cf. Ref.\cite{Blatter} for a more detailed discussion of this subject).

Recently, it was noted that granular superconductors may become glassy even
in the absence of external magnetic field, if a large enough part of Josephson
junctions are  anomalous, i.e. their minimum Josephson coupling
energy corresponds to a phase difference $\Delta \phi = \pi$
instead of $0$ (so-called $\pi$-junctions). Two completely different
origins of $\pi$-junctions were proposed: mesoscopic fluctuations in dirty
superconductors \cite{spivak} and the pairing with non-zero momentum
\cite{gesh,sigrist}. Recent experiments revealing the $d$-wave nature
of pairing in high-temperature superconductors \cite{d-wave} indicate
the possibility of observing glassy superconductive behaviour in HTSC ceramics
in virtually zero magnetic field. Note that ceramics with equal
concentrations of usual and $\pi$-junctions are completely equivalent
(if screening effects can be neglected) to the $XY$ spin glass. 
Contrary to the 3D gauge glass model, the XY spin glass in 3D is expected to
have no true thermodynamic phase transition at finite temperature \cite{huse}; 
 recently, it has been suggested that the XY spin-glass and d-wave 
ceramic superconductor might have a new equilibrium ordered phase, the 
so-called chiral-glass phase \cite{kawamura}. However,
these issues are hardly relevant for the measurable response at temperatures
much below ``bare'' glass transition temperature $T_g$, which we consider in
 this paper.

Experimental studies of granular superconductors reveal \cite{Vela,Morgen}
an appearance of
magnetic irreversibility (a difference between Meissner and
shielding magnetizations
or, in other terms, between Field Cooled (F.C.) and Zero Field Cooled (Z.F.C.)
magnetizations)
below some temperature $T_g$, which is lower than the SC transition
temperature $T_c$ of the grains. However, detailed analysis of the magnetic
response in such systems is usually complicated by the mixing of contributions
from individual grains and from the intergrain currents.
The  goal of this
paper is to develop a method which makes it possible to
extract from the raw data
on d.c. magnetic response the intergrain contribution and to
compare its behaviour with existing theoretical predictions.

The compound La$_{1.8}$Sr$_{0.2}$CuO$_4$~ was chosen in this study for experimental
convenience, since its critical temperature ($\approx 32\,K\,$) is within
the optimal temperature range of our noise and a.c. susceptibility
measurements setup.
The sample was fabricated by standard solid state reaction of
La$_2$O$_3$, SrCO$_3$ and CuO \cite{Jehan}. Mixed powder was pressed into  pellets which were sintered
in air at $920^\circ$C for 12 hours. The material was then submitted
to three cycles of regrinding, sifting to $20\, \mu m$, pressing and
sintering  again at $1100^\circ$C for 12 hours. Samples prepared
in two successive runs were used in this study. In the first
one (sample A), pellets $1\, mm$ thick and $10\, mm$ diameter were
obtained, with a density about $80\%$ of the theoretical bulk value.
In the second one (sample B), cylinders of diameter $6\, mm$ and length
5 to $6\, mm$ were prepared with a density ratio about $88\%$.
In both preparations, grains sizes were in the range 1 -- 10 $\mu m$. Room
temperature X-ray powder diffraction patterns showed the presence of a
small amount ($ < 5\%$) of the non superconductive compound
La$_{1 - 2x}$Sr$_{2x}$Cu$_2$O$_5$.

The rest of the paper is organized as follows. In Section II
the general analysis of the
magnetic response data obtained on two different samples
(A and B) of La$_{1.8}$Sr$_{0.2}$CuO$_4$~ ceramics is presented
and the intergrain (Josephson) contribution to the overall response
is extracted. Section III is devoted to the detailed study of the magnetic
response of Josephson intergrain network in the low-field
range. It is found that the macroscopic critical current is suppressed
considerably (by a factor 2), in a magnetic field of only about $2G$.
The lower-field d.c.--response to field variations of order
$0.05 \div 0.5 G$
was analyzed for the F.C. states obtained at $H_{FC} = 0 \div 10 G$
and two temperatures, $10K$ and $20K$.
The data at $T=10 K$ and $H_{FC} = 0$ and $0.1 G$ are
shown to be compatible with the Bean critical-state picture \cite{Bean}
and the low-field
critical current value is identified. The rest of the data are in a sharp
contrast with Bean-model predictions: the screening current
{\it grows sublinearly} (approximately as a square root) with increasing  exitation field. Very low field, low frequency a.c. measurements
are presented, which reveal the strongly irreversible nature of that anomalous
response. A new phenomenological model is proposed for the treatment of
these data. Its first predictions are found to be in a reasonable agreement
with the data. In Section IV, the theoretical analysis of our experimental results
is given in terms of the existing theories of ``gauge-glass'' state. It is shown
that  the observed transition temperature to the low-temperature state of the
network and the magnitude of the (low-$B,T$) critical current are in sharp contradiction
with the (usual) assumption that the zero-field granular network is unfrustrated.
On the contrary, under the assumption of a strongly frustrated network 
at $B=0$, all basic measured parameters of the ceramic network are in mutual
agreement. We believe that these estimates indicate the existence of a large
proportion of $\pi$-junctions in the La$_{1.8}$Sr$_{0.2}$CuO$_4$~ ceramics, possibly due to the $d$-wave
nature of superconductivity in cuprates. The Section V is devoted to the
development of a new model of diamagnetic response in glassy superconductors,
which is necessary for the description of the anomalous data described at the
end of Section III. This new model (in some sense, intermediate between
the Bean \cite{Bean} and the Campbell \cite{Campbell} ones) is based on two ideas:
i) the existence of two characteristic ``critical'' currents ($J_{c1}$ and
$J_c \gg J_{c1}$), and ii) the fractal nature of free energy valleys in the ceramic
network. Our conclusions are presented in Section VI, whereas some
technical calculations can be found in the Appendix. 

For convenience, the e.m.u system of units will be used for 
experimental data, and Gaussian units for the theoretical discussions.

\section{General Properties of D.C. Magnetic Response}
      The d.c. magnetization was measured by the classical extraction 
method. Two
      SQUID magnetometers were used: one
      a home made apparatus used in several previous spin-glass
      studies \cite{Alba}, the other a commercial system (Cryogenics S500).

      In this section, we describe successively the static magnetic
      response of samples A and B and present a preliminary treatment of
      these data, in order to distinguish between the magnetic response of
      individual grains and intergrain currents \cite{Vela,Dersh} (a detailed
      study of the latter is the subject of the next section). Firstly, we 
      present results obtained after cooling the samples in various d.c. fields
      and applying small field increases.
      Secondly, we will derive from the results the response of the Josephson
      currents as a function of field and temperature. Finally, we will show that
      the behavior of the field cooled (F.C.) susceptibility can be satisfactorily
      accounted for if the system of Josephson currents does not carry Meissner
      magnetization. It will be shown that the same interpretation accounts 
      fairly well
      for the F.C. results which, at first sight, are rather different
      for the samples A and B. 
      \subsection{Sample A}
      Sample A is a $1\,mm$ thick pellet with an approximately ellipsoidal
      shape of $2\times 6\,mm$. Its calculated volume is $V\approx 8.5\,mm^3$
      and the demagnetizing field coefficient for the field parallel to the
      longitudinal axis is $N\approx 0.06$ \cite{Demag}.\\

\begin{figure}[htb]
\epsfxsize=\hsize \epsfbox[0 0 400 300]{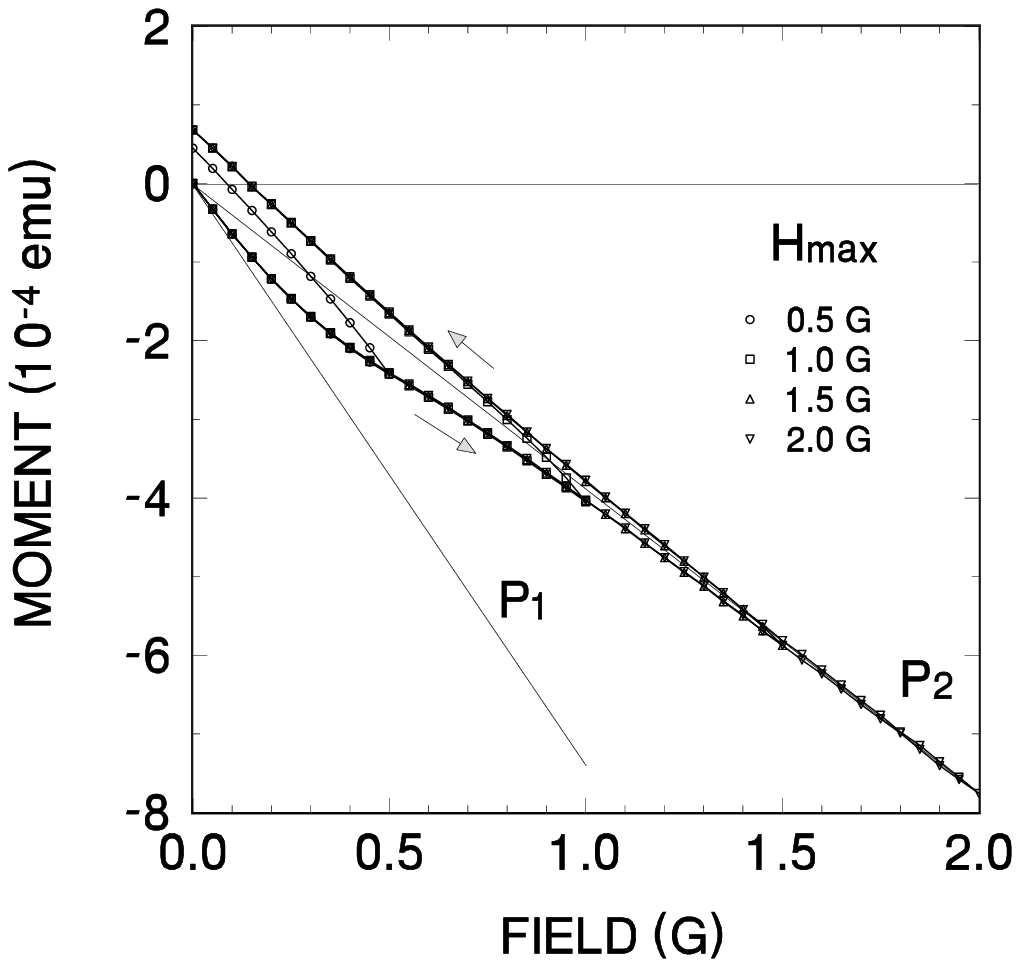}
\caption{Magnetic moment of the sample A as a function of field applied in 
the zero field cooled state (e.m.u. units of moment correspond to 
$cm^3-G$).}
\label{ahyst}
\end{figure} 

Fig.~\ref{ahyst} displays the magnetic dipole moment of the sample
      cooled to $10\,K$ in zero field and submitted to cycles
      $0 \rightarrow H_{max} \rightarrow 0$ for several
      values of $H_{max}$ up to $2\,G$. At the lowest increasing fields, the
      moment increases initially with a slope $P_1$. Above $1.5\,G$, 
it approaches
      a slope $P2$. The remanent positive moment saturates for $H_{max}\geq 1\,G$.
      The calculated moment of the sample for perfect volume shielding
      in an homogeneous field is (e.m.u. system):
      $${\cal M} = -{{H\cdot V}\over {4\pi (1 - N)}}
      = -0.72\cdot 10^{-3} \times H\, cm^3{\mbox{\rm -}}G\,.$$
      Owing to the error in the evaluation of the volume, this value is determined
      with an accuracy of only $\pm 5\%\,$. Nevertheless, it is
      in fair agreement with the slope $P1$ in Fig.~\ref{ahyst}.
      On the other hand, the  slope $P2$ is about $53\%$, a rather small value
      since the density ratio of the sample is about $80\%$. At such low
      temperatures (in comparison with $T_c \approx 32\,K$), where the lower critical field
      of the grain's material is
      above $100\,G$, one would expect expulsion of the field by the grains
      with a penetration depth $\lambda$. The expected value for the magnetization
      $M = {\cal M }/V$ of the system of uncoupled grains system  can be calculated 
      as \cite{Yaron}:
      \begin{equation}
      {M\over H}={1\over 4\pi}\cdot {f\over {1 - fN - (1-f)n}}
      \label{eq-yaron}
      \end{equation}
      where $f$ is the volume fraction of the superconductive material and
      $n$ is the demagnetizing field for the grains.
      For an estimate, we assume grains to be spherical
 ($n=1/3$) and, using $M/H \approx 0.53\cdot 1/4\pi$ and $N=0.06$, we find
  $f \approx 0.41$. This value is considerably below the volume fraction
  of the sample filled by grains ($\approx 0.8$); we assume that the difference
  is due to
  the intergrain penetration depth $\lambda$ being comparable to the
  grain size $r$ and estimate an effective value of $\lambda$ as
      $$f=0.41=0.8\, \left(1 - {\lambda\over r}\right)^3
      \mbox{\rm ~~~~~yielding~~~~~}
      \lambda = 0.2\,r\,.$$
      Taking an average size of $5\,\mu m$ for the grains, we obtain $\lambda
      \approx 500\,nm$. Values reported for the mean penetration depth
      in La$_{1.8}$Sr$_{0.2}$CuO$_4$\ are about $200\,nm$ \cite{Poole}. The value found here is
      larger than the expected mean value for the homogeneous material,
      indicating that the grains are not monocrystalline. This will be confirmed
      below by the results of field-cooling experiments.

\begin{figure}[htb]
\epsfxsize=\hsize \epsfbox[0 0 400 300]{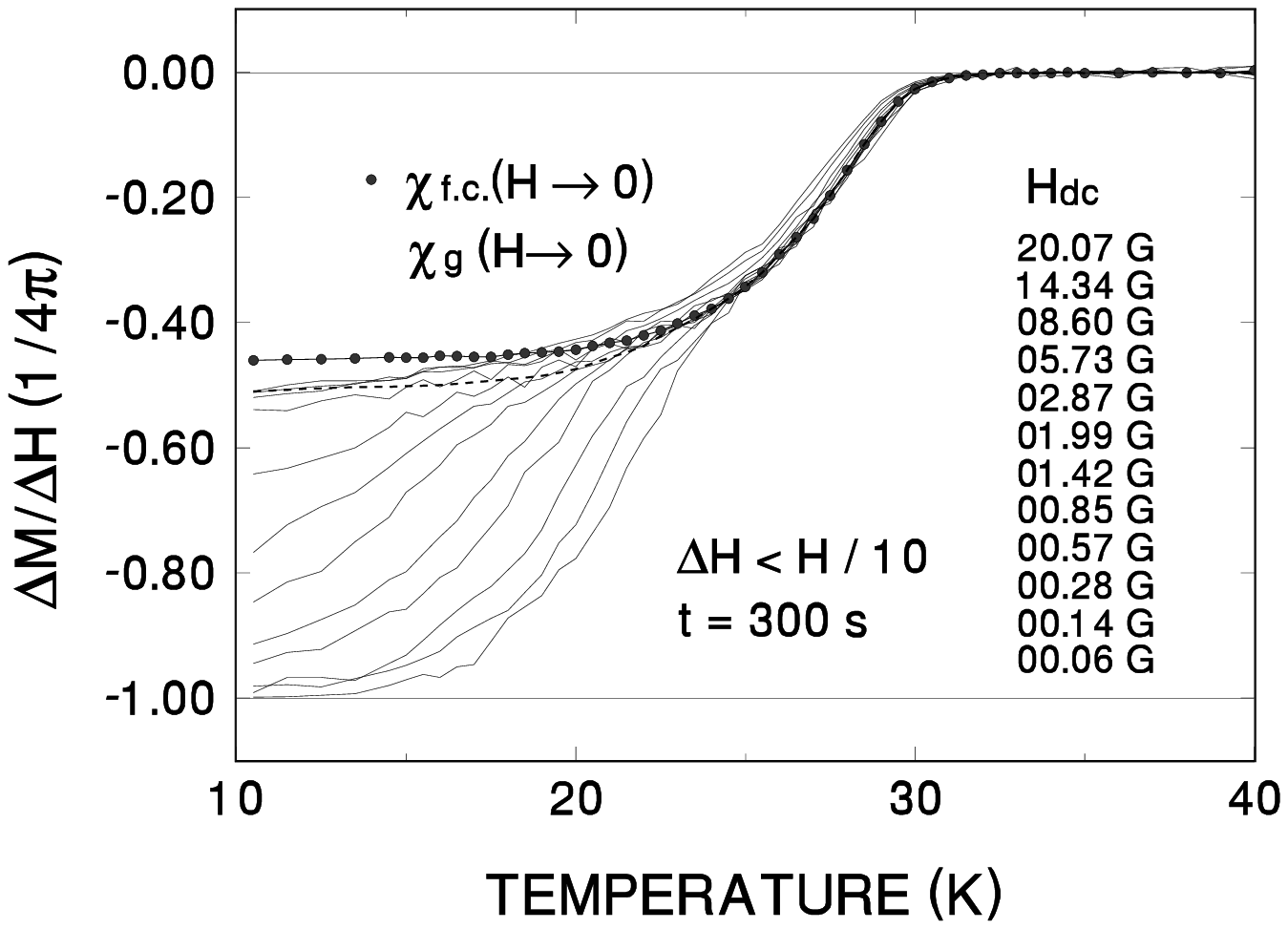}
\caption{Shielding susceptibiliy of the sample A as a function of temperature,  
normalized to the moment for complete shielding. Curves are arranged in the same ascending order as in the legend.}
\label{amdhvst}
\end{figure}

      The shielding susceptibility is plotted in Fig.~\ref{amdhvst}, as a
      function of temperature and for several values of the ambient F.C. field.
      The measurements were performed according to the following procedure:
      the sample was cooled in a field $H_{dc}$ down to the working temperature
      and the moment was measured after waiting 300 sec; then the field
      was increased by a small amount $\Delta H \leq H_{dc} /10$ and the
      moment was measured again after waiting 300 sec. The figure displays
      the experimental shielding susceptibility normalized to the value for
      total shielding, i.e.:
      $$\chi_{sh}={{{\cal M}(H+\Delta H) -{\cal M} (H)} \over \Delta H}
      \times {{4\pi(1-N)}\over V}\,.$$
      The curves show the double step usually ascribed to the action of
      both intragrain currents and Josephson intergrain currents \cite{Renker}.
      At high temperature, the onset of grains diamagnetism occurs at about $32\,K$.
      Above $25\,K$, the response corresponds to the diamagnetism of the grains.
      At a fixed temperature, it is $H_{dc}$ independent for $H_{dc} \leq 5\,G$,
      and decreases for increasing
      $H_{dc} > 5\,G$. Below $25\,K$, the onset of Josephson currents manifests as
      a second step of the diamagnetic response. This second step appears at
      a decreasing temperature as $H_{dc}$ increases. At the lowest temperatures,
      the diamagnetic moment amounts to about $100\%$ of flux expulsion at
      $H_{dc}=0$ and decreases with increasing $H_{dc}$. At $H_{dc} > 8\,G$, the
      flux expulsion saturates at a value slightly above $50\%$ which corresponds
      roughly to the level of $53\%$ determined above for the grains response.

      The susceptibility in Fig.~\ref{amdhvst} contains the contributions
      of grains and Josephson currents. The contributions can be separated
      on the line of the work by Dersh and Blatter \cite{Dersh}. The induction
      in the sample is given by 
      $B = H+4\pi(M_g +M_j )$
where $M_g$ and $M_j$ stand respectively for the magnetization of grains and 
of the Josephson currents.       
It should be noted that the magnetization due to macroscopic circulating currents
in a superconductor is sample-size dependent, i.e. the corresponding
susceptibility is not a local quantity.
      At the macroscopic scale of the circulating currents, the
      magnetization $M_g$ can always be written as $\chi_g H_{local}$, where
      $\chi_g (H)$ is homogeneous over the sample.
      In what follows, we consider quantities averaged over the volume of the
      sample: in that case, $M_j$ is the averaged moment per volume unit due to the
      currents.
      The demagnetizing field effect will be neglected in the calculations. We have
      verified that, owing to the small value of the demagnetizing factor ,
      this does not modify the essential features of the result while allowing
      a simpler derivation (the effect of demagnetizing factor will be taken
       into
      account when analyzing the data from the  sample B). We get
      $M_g = \chi_g (H+4\pi M_j )\,.$
      Then
      $M = M_g + M_j = \chi_g H+M_j \mu_g$ (with $\mu_g = 1+4\pi\chi_g\,$),
      and
      \begin{equation}
      M_j = {{M-\chi_g H}\over \mu_g}\,.
      \label{eq-m_j}
      \end {equation}
      Eq.~(\ref{eq-m_j}) must be considered with care since $\chi_g$ is history
      and field dependent. In fact it is well-adapted to the description of the
      result of zero (or small) field cooling experiments. More generally, we must consider
      the response to field increments $\delta H$ to obtain 
      $\chi = {\delta M / \delta H}\,.$ 
      Then, the polarizability \cite{note} $\chi_j$ 
      of the Josephson network reads:
      \begin{equation}
      \chi_j = {{\chi - \chi_g}\over \mu_g}\,.
      \label{eq-chi_j}
      \end{equation}
      Note that we can equivalently consider the response of the currents system
      in an homogeneous medium with permeability $\mu_g$. If the applied field is
      varied by $\delta H$, the Josephson network sees a variation of internal field
      $\delta H_i = \mu_g \delta H$ and develops a polarization
      $\delta M_j = \chi_j \delta H_i\,$. Then, we recover Eq.~(\ref{eq-chi_j}).

\begin{figure}[htb]
\epsfxsize=\hsize \epsfbox[0 0 400 300]{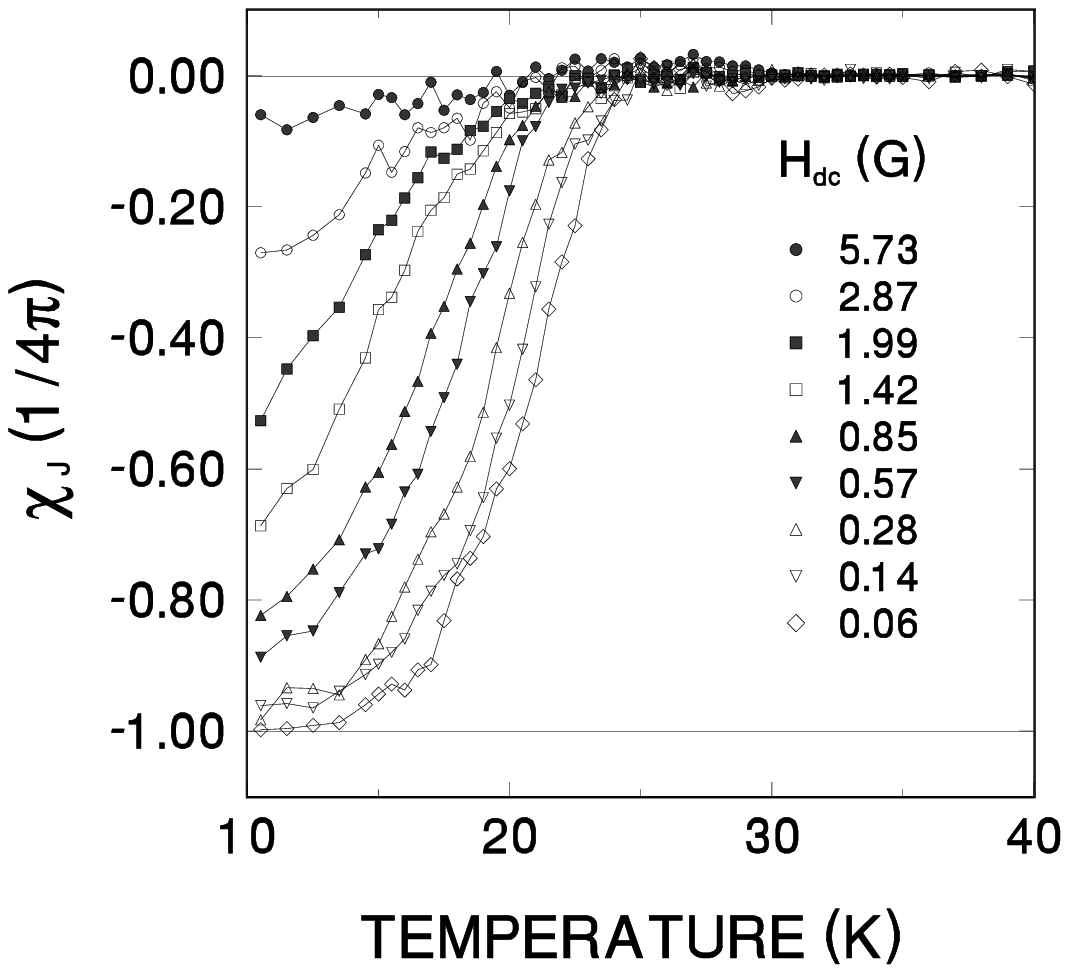}
\caption{Josephson currents susceptibility in the sample A as a function of 
temperature. Values have been calculated from data of Fig. 
\protect\ref{amdhvst} and using Eq. 
(\protect\ref{eq-chi_j}).}
\label{akhijdc}
\end{figure}

      The value of $\chi_g$ could be determined in principle if we were able to
      obtain a packing of disconnected grains equivalent to the packing of the
      sintered sample.
      In practice this was not possible. Indeed, mechanical grinding resulted
      in breaking a large part of the grains and thus modifying the characteristics
      of the material. Nevertheless, it is possible to extract $\chi_g$, at least
      approximately, from the data of Fig.~\ref{amdhvst}. At high temperature,
      above the onset of intergrain currents at $\approx 25\,K$, the shielding
      susceptibility $\chi_{sh}$ is due to the grains alone,
      independent on $H_{dc}$ below
      $\approx 6\,G\,$. At low temperatures, for $H_{dc}$ above $\approx 6\,G\,$,
      the $\chi_{sh}$ curves superpose and there is no manifestation of
      the onset of intergrain currents. Thus , here also, 
      $\chi_{sh}$ represents the
      response of the grains alone. Hence, the response $\chi_g$ of the grains
      can be reasonnably approximated by an interpolation
      between these two limits. The interpolation curve, obtained by a smoothing
      procedure between both curves at $H_{dc} = 0\,G$ and $H_{dc} = 20\,G$ is
      displayed on Fig.~\ref{amdhvst} (dotted curve). The
      values of $\chi_j$ derived
      from Eq.~(\ref{eq-chi_j}) are plotted versus temperature in Fig.~\ref{akhijdc}, for $H_{dc}<6\,G$.

      Note that the dependence of $\chi_j$ on $H_{dc}\,,$ seen in
      the figure is supposed to reflect the behavior of the initial shielding
      properties of the Josephson network with the increase of $H_{dc}$. Nevertheless,
      non linearity of the response due to the correlative increase of the value
      of $\Delta H$ ($\Delta H = H/10$) cannot be excluded: this aspect will be
      studied in detail
      in sample B. Finally, one can note the   similarity of our data with the
      results of earlier numeric simulations on a gauge glass system \cite{Morgen}.

\begin{figure}[htb]
\epsfxsize=\hsize \epsfbox[0 0 400 300]{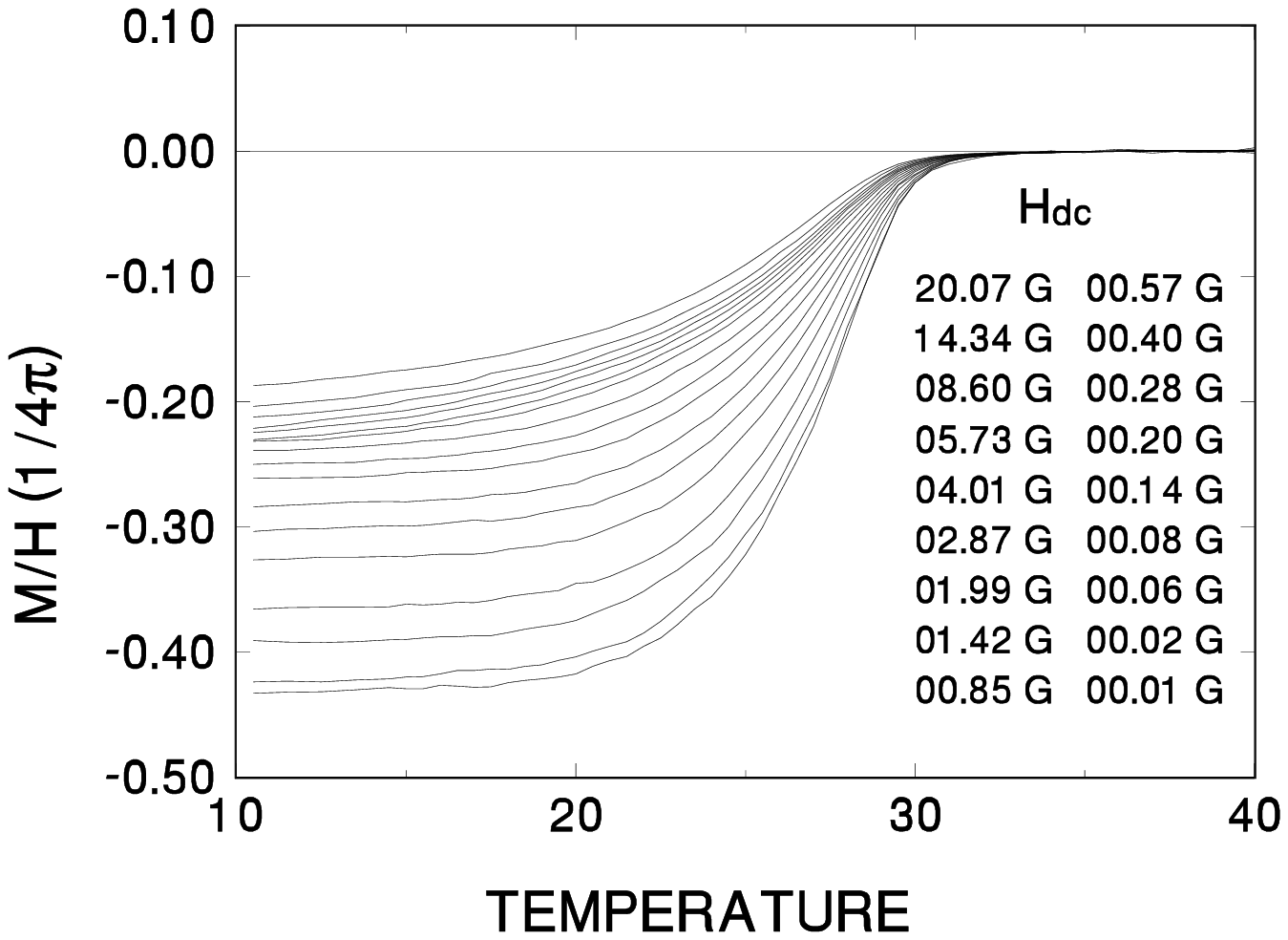}
\caption{F.C. (Meissner) susceptibility of sample A as a function of 
temperature for fields up to $20\,G\,$. Curves are arranged in the same 
ascending order as in the legend.}
\label{amfcvst}
\end{figure}

\begin{figure}[htb]
\epsfxsize=\hsize \epsfbox[0 0 400 300]{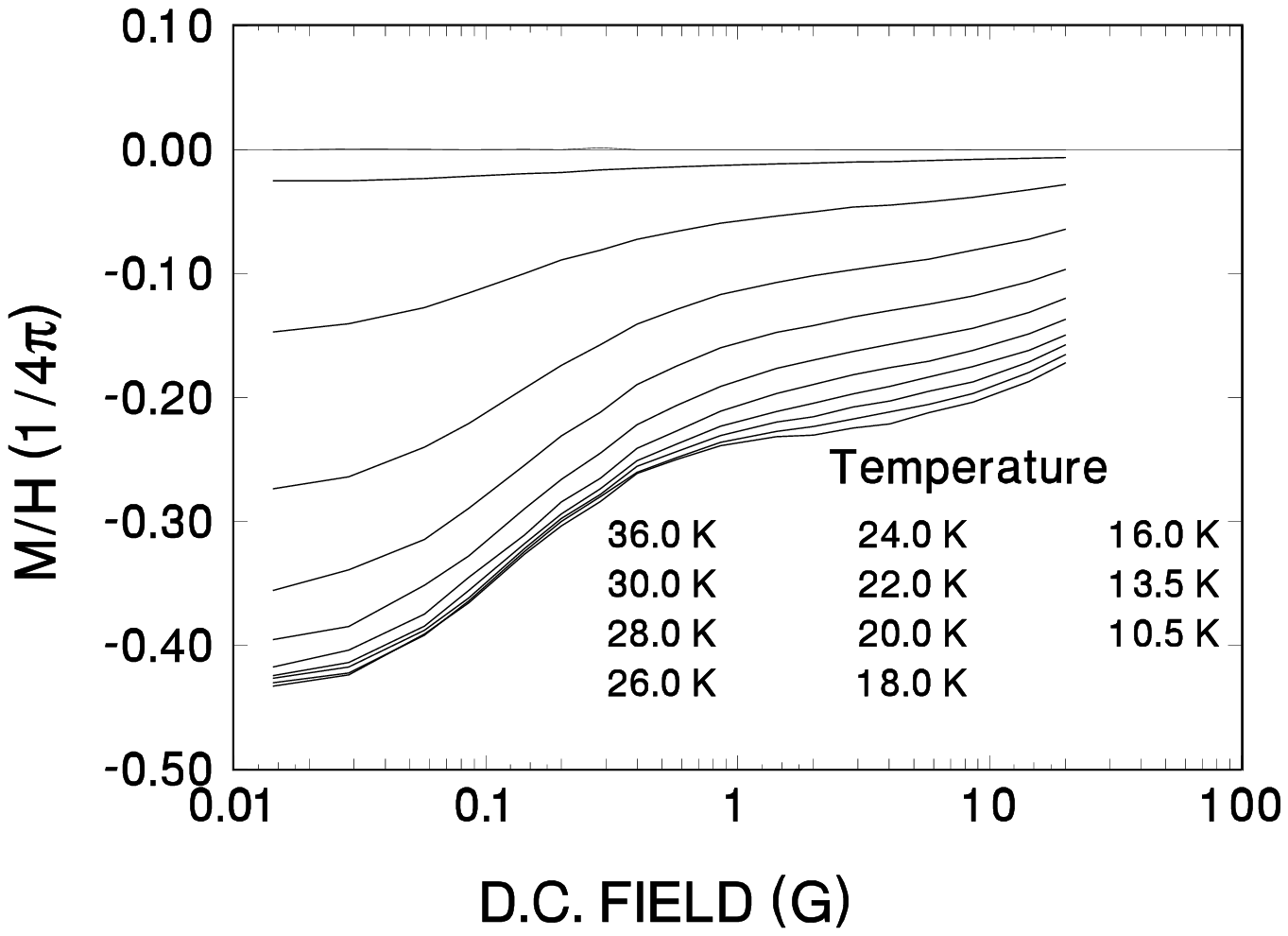}
\caption{F.C. (Meissner) susceptibility of sample A as a function of 
field for selected temperatures Curves are arranged in the same ascending 
order as in tthe legend.}
\label{amfcvsh}
\end{figure}

Above we have dicussed the system's responses to the variation of magnetic
field at fixed temperature (i.e. shielding responses) and extracted 
from these
data the polarizability $\chi_j$ of the intergrain system.
Now we turn to the
description of the results of
the Field Cooling (F.C.) measurements.
      F.C. (Meissner) magnetization was measured by the standard procedure between
      $10K\,$ and $40\,K$ for fields from $0.01$ to $20\,G$. The results are
      reported in Fig.~\ref{amfcvst} versus temperature and Fig.~\ref{amfcvsh}
      versus applied field. Data are normalized to the value of the moment for
      $100\%$ shielding.
     
      Even at the smallest field, the flux expulsion rate is no more than
      $45\%$, less than the $53\%$ shielding by the grains. At low fields, below
      $1\,G$, there is an approximate affinity between the curves of
      $M/H$ versus $T$.
      $M/H$ can be extrapolated linearly to $H \rightarrow 0$. The result is
      plotted in Fig.~\ref{amdhvst} (solid circles): one can see that the
      extrapolated F.C. susceptibility superposes exactly with the low d.c.
      field shielding susceptibility above $25\,K$.
      Therefore, at low d.c. field above $25\,K$, the response of the grains system
      is reversible and it is well described by the low d.c. field shielding
      curves; this justifies the hypothesis  used above for the calculation
      of $\chi_j\,$.
      On the other hand (see Fig.~\ref{amfcvsh}), the behavior of the F.C.
      susceptibility as a function of $H$ is not trivial. $M/H$ decreases with
      increasing field and reaches a stable level
      (about $25\%$ at the lowest temperatures)
      at roughly $1\,G$. Whatever the  temperature,
      this decrease is centered at a constant value of the field, about
      $0.1 - 0.3G$. Above  $5G\,$, $M/H$ decreases once more with
      increasing field.
Note an essential difference between the
F.C. results presented on Fig.~\ref{amfcvst}
and the shielding results above
      (Fig.~\ref{amdhvst}): the F.C. curves do not show any increase
      of the response $M/H$ with the temperature decrease below 20K, where
      the intergrain coupling grows considerably (as it is seen from
      Fig.~\ref{amdhvst}). This means that the network of intergrain
      currents does not produce Meissner (F.C.) magnetization, whereas
      it does produce {\it shielding} magnetization.

      The behavior of the F.C. susceptibility $\chi_{FC} = M_{FC}/H$ as 
      a function of the
      applied field $H$  depicted in Fig.~\ref{amfcvsh}
      shows two nontrivial features: i) crossover between two plateaus (at low and
      moderate values of $H$), which takes place between $0.1 G$ and $1 G$
      independent of temperature, and ii) the value of the low-field $\chi_{FC}$
      is noticeably lower than the Meissner response of uncoupled
      grains ($45\%$ versus $53\%$).  These features
      can be understood in terms of (i), a polycrystalline structure of the grains,
      which can be suspected from the large values of the penetration depth
      obtained from the results of Fig.~\ref{ahyst},
      and (ii), self-shielding (pinning of the magnetic flux)
      by the Josephson currents when lowering the temperature in an
      applied field.

We start from the feature i);   the curves of F.C. magnetization in
Fig.~\ref{amfcvsh} are rather
      similar to those which were measured by Ruppel et al.
      \cite{Ruppel} in YBaCuO ceramics. The authors interpreted their results
      on the basis of a theory
      of the flux expulsion by strongly anisotopic randomly oriented crystallites
      as derived by Wohllebeen et al. \cite{Wohll}. We stress that the
      model is not based on any activated flux creep mechanism. It is thus 
well-adapted to the analysis of our results: indeed, flux creep effects can
      hardly be invoked here since the temperature
      has no apparent effect of on the characteristic field
      related to the decrease of magnetization. The starting point of the model
      is that, provided the size $b$ of the crystallites is such that
      $\lambda_{\parallel} \ll b \ll \lambda_{\perp}$, the longitudinal
      magnetization of a crystallite whose c-axis makes an angle $\alpha$
      with the field is given by  $M = -(H/4\pi)\cdot \gamma (\cos \alpha)^2$,
      where $\gamma$ is a factor close to 1, depending on the ratio
      $\lambda_{\parallel} / b$. After averaging over $\alpha$, one obtains
      $M/H = (\gamma /3)(1/4\pi)$.
      It must be stressed that the system of intragrain crystallites 
      is a strongly-coupled system, contrary to the system of grains which
      composes the ceramic. Therefore, a grain consists of an ensemble of
      interconnected Josephson loops surrounding crystallites whose planes are
      nearly along the field and are thus transparent to the field. At low
      fields, this system will expel the flux with a penetration depth
      depending on the junction coupling energy. Nevertheless, when the
      field is such that a loop sees a flux larger than $\sim\Phi_o /2$, the
      macroscopic magnetization of the Josephson currents vanishes and the
      system reacts as an ensemble of disconnected crystallites \cite{Stroud}.
      The characteristic field of this crossover is such that \cite{Wohll}:
      \begin{equation}
      {{H_m s_c}\over \Phi_o} \approx 0.1
      \label{eq-wohll}
      \end{equation}
      Recently determined values for the penetration depth in La$_{1.8}$Sr$_{0.2}$CuO$_4$\ 
      \cite{shiba} are $\lambda_{\parallel} = 150\,nm$ and $\lambda_{\perp}
      = 1500\,nm$. Older measurements indicate a higher anisotropy,
      up to a factor 14 \cite{Suzuki}. We can thus reasonnably consider that
      the model can be applied in our case. Taking $H_m = 0.3\,G\,$, we obtain
      $s_c = 7.4\cdot 10^{-8} cm^2\,$. With $s_c \approx \pi b^2$ this leads to
      a mean diameter $b = 1.5\,\mu m$ for the crystallites.
      Above $H_m$, the system acts as
      an ensemble of crystallites whose average susceptibility is
      $(\gamma /3)(1/4\pi)\,$. With the density ratio $f = 0.8\,$, taking
      $\gamma =1$ and supposing
      spherical crystallites we obtain from Eq.~(\ref{eq-yaron})
      $4\pi{M\over H} = 0.31$ which is above the experimental value
      ( the latter being about $0.25$). Nevertheless, it must be noted
      that we have neglected here the effect of the factor $\gamma$ and used
      a rather unrealistic spherical approximation for the shape of
       crystallites.
      Finally, it has been seen that above $5\,G\,$, the F.C. magnetization
      starts to decrease once more with increasing field although $H_{c1}$ is
      larger than $100G$ in La$_{1.8}$Sr$_{0.2}$CuO$_4$\, . This can be due to intrinsic pinning inside
      the crystallites themselves when the applied field is such that the flux
      in the cross section of one crystallite is larger than $\Phi_o\,$. With a
      mean radius of $0.8\,\mu m$ for the crystallites, this crossover occurs at
      about $10\,G\,$.

\begin{figure}[htb]
\epsfxsize=\hsize \epsfbox[0 0 400 300]{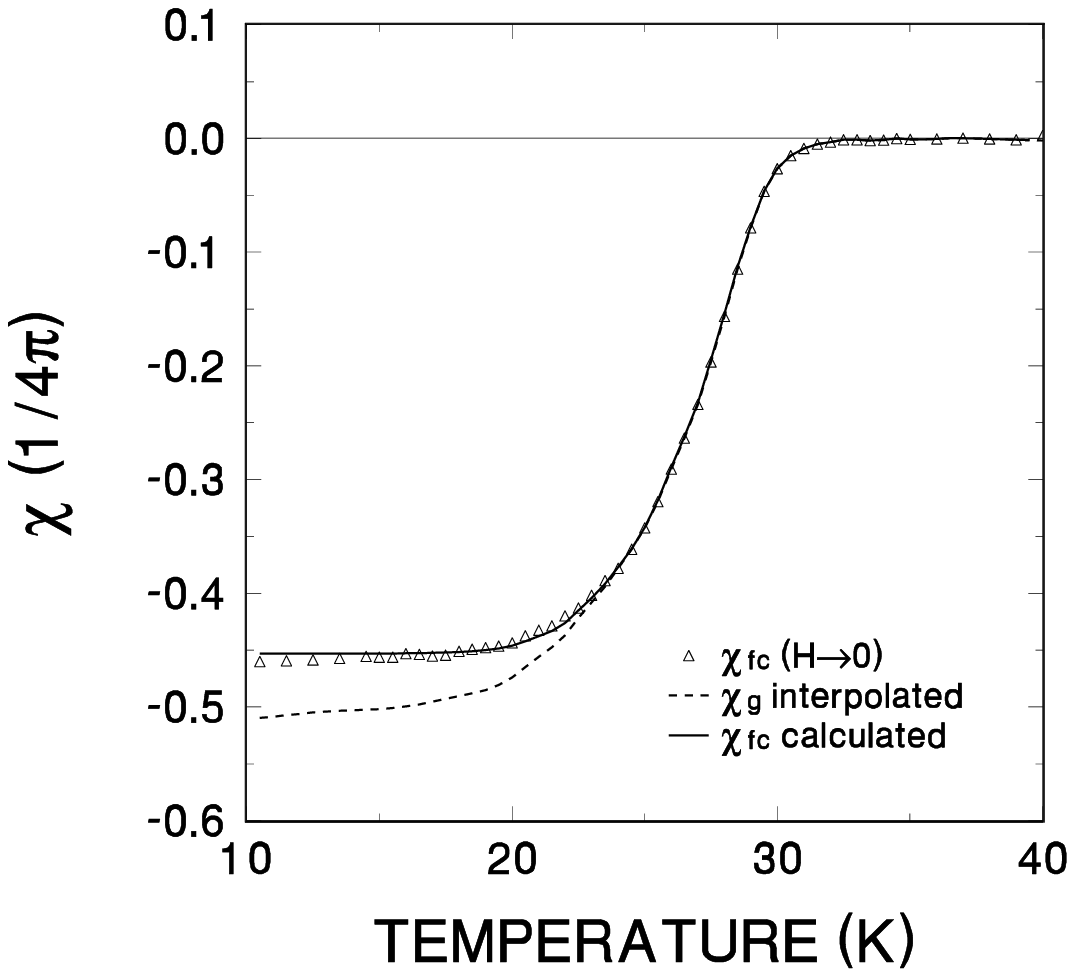}
\caption{$\chi_{FC}$ calculated with Eq.~(\protect\ref{eq-chi_fc}) from the 
values of $\chi_g$ and $\chi_j (0)$ (see text).}
\label{afccalc}
\end{figure}

Now we turn to the discussion of the feature ii) mentioned above.
At temperatures below $25\,K\,$, the Josephson currents become active. 
Their effect is that,
      at $10\,K\,$, the shielding response of the system of grains amounts at
      about $53\%$, while the
      F.C. susceptibility saturates at about $45\%$. This difference is enough
      to be significant and can be interpreted as the result of pinning by the
      Josephson network. In fact, this pinning can be
      understood as a back shielding effect of the Josephson currents against the
      decrease of local internal field, due to the temperature dependence of the
      grain's system permeability  $\mu_g$. We have seen
      above that the response of the system consists of the two parts: (i) for an
      applied field $H$, the internal field due to the grains seen by the
      intergrain currents is $H_i = \mu_g H\,$, and (ii) the intergrain currents
      system reacts to all variation of $H_i$ with a polarizability $\chi_j$
      and generates a magnetization $\delta M_j = \chi_j \delta H_i$. Thus, when
      the temperature is decreased by $dT$, the internal field decreases by
      $H\,d\mu_g /dT$ and the Josephson network tends to screen this variation.
      Since the intergrain currents give no Meissner effect, we consider their
      response as totally irreversible. Thus
      for a variation $dT$ of the temperature, in a field $H\,$, the variation
      of induction is:
      $$dB = (1 + 4\pi \chi_j )\left({{d\mu_g }\over dT}\right)_H H\, dT$$
      On the other hand,
      $B = (1 + 4\pi \chi_{FC})\,H\,$.
      With $\mu_g = 1 + 4\pi \chi_g\,$, we
      finally obtain:
      \begin{equation}
      \chi_{FC} = \chi_g + 4\pi \int^T _{T_c} \chi_j\, {d\chi_g \over dT}\, dT
      = \chi_g + \chi_j^{FC}\,.
      \label{eq-chi_fc}
      \end{equation}
      $M_j^{FC}=\chi_j^{FC}\,H$ is the magnetization produced by
       the Josephson currents
      due to  variation of $\mu_g$ with decreasing temperature.
      As $\chi_g$ is known only in the limit $H_{dc}\rightarrow 0\,$,
      Eq.~(\ref{eq-chi_fc}) has been used to calculate $\chi_{FC}$ versus
      $T$ in the limit of low field. In order to do it, we started from the values of
      $\chi_g (H\rightarrow 0)$ as derived above; for $\chi_j$, we have used the values
      given in Fig.~\ref{akhijdc} for the smallest field $H_{dc} =0.06\,G\,$.
      The result is plotted on Fig.~\ref{afccalc}. The agreement of calculated
      values with experimental data is rather satisfactory, although not perfect.

This discrepancy is emphasized if Eq.~(\ref{eq-chi_fc}) is
      reversed in order to calculate $\chi_j$ as a function of $\chi_g$ and
      $\chi_{FC}\,$. The reason is that we have used here the simplest linear
      model of
      back shielding. In fact, as we will see later, the response of
      the currents system is strongly non-linear, with the susceptibility $\chi_j$
      decreasing
      with increasing $\Delta H\,$, and this effect becomes stronger as the
      temperature increases.
      The result is that the calculated efficiency of back shielding is
      underestimated, since the value of the experimental susceptibility is
      determined by applying finite increments $\Delta H\,$.

\subsection{Sample B}
      Sample B was machined from one of the original cylinders, in form of a
      parallelepiped of dimensions approximately $3\times 3\times 6\,mm^3\,$.
      Its calculated volume is $V\approx 52.6\,mm^3$ and its demagnetizing
      field coefficient for a longitudinal field is $N\approx 0.19\,$. In a
      longitudinal field, its calculated moment for perfect flux expulsion
      is given by
     ${\cal M} = 5.1\pm 0.2\cdot 10^{-3} \times H \> cm^3{\mbox{\rm -}}G$

      Measurements of the initial magnetization at $10\,K$ are in fair 
agreement with this value. For $H_{dc}$ above $3\,G$ and up to $30\,G\,$
      $\Delta {\cal M} /\Delta H$ reaches a stable
      level about $3.2\,10^{-3}\,cm^3\,$ which corresponds to the response of the grains alone. With the density ratio
      of $88\%$ for this sample and using Eq.~(\ref{eq-yaron}) one finds
      $f = 0.46\,$, yielding $\lambda = 0.19\,r\,$,
      i.e. the same value as derived for sample A.

      The shielding susceptibility was measured in this sample by using a
      more sophisticated method, in order to reduce the effect of non linearity.
      After cooling the sample at the working temperature in the d.c. field, the
      field was increased by 5 successive steps $\Delta H\,$, and $\Delta\cal M$
      was measured. At the lowest
      fields, $\Delta H = 10\,mG$ and (to keep a good signal/noise ratio)
      $\Delta H = H_{dc}/50$ at the highest ones. Then, the value of
      $\Delta{\cal M}_n/\sum_n\Delta H$ was extrapolated to $\Delta H = 0$ by least
      square fit.  

      Like in the case of sample A, all
      curves at $H_{dc} \leq 10\,G$ merge at high temperatures to a common
      curve which corresponds to the flux expulsion by the grains.
      The main difference with the sample A is that in the sample B the onset of
      Josephson currents shielding
      occurs at higher temperatures. This is consistent with the fact that
      sample B is more dense, resulting in a better coupling between grains;
moreover, its size is larger, which also increases the total shielding magnetization.  At low temperature, the magnetization
      curve at $H_{dc} = 20\,G$ reaches a level slightly
      above $60\%$, which corresponds to the low temperature level for the
      grains.

\begin{figure}[htb]
\epsfxsize=\hsize \epsfbox[0 0 400 300]{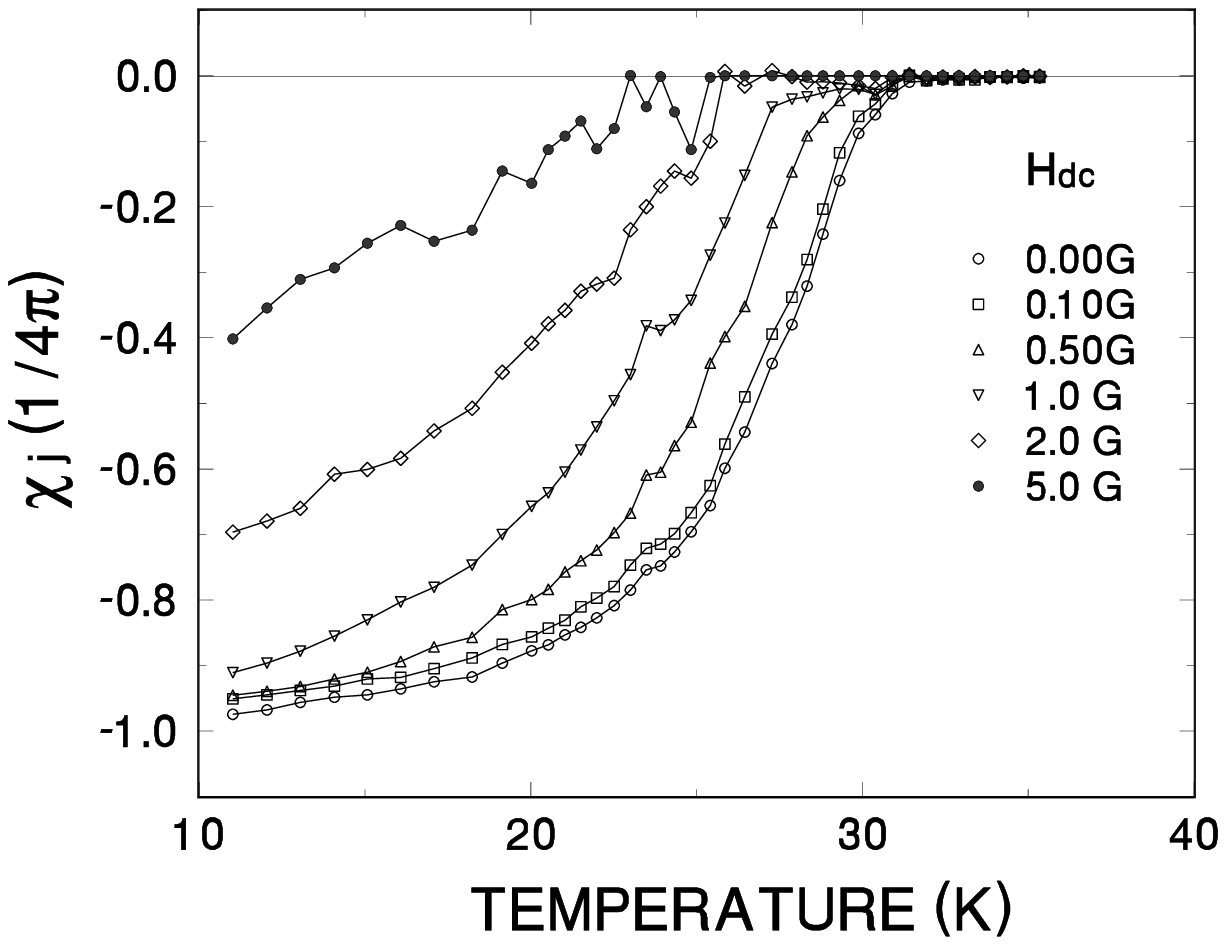}
\caption{Josephson currents shielding susceptibility as derived from 
the data and use of 
Eq.~(\protect\ref{eq-chi_j2}).}
\label{bkhijic}
\end{figure}

      The shielding response of the Josephson currents is obtained with 
      the procedure already used for the
      sample A. Here the demagnetizing factor cannot be neglected
      ($N\approx 0.19\,$). Two kind of quantities are to be considered:
      (i) the responses $\chi_g$ and $\chi$ of an equivalent sample
      without
      demagnetizing field (e.g. an infinitely long cylinder with the
      same cross-section for instance); here
      $\chi_g$ is the response of the system of grains
      alone, without intergrain currents, and $\chi$ is the total
      response of the system of
      intragrain plus intergrain currents,
      and (ii), the measured responses
      $\overline{\chi}_g$ and $\overline{\chi}\,$
      ; they correspond to the
      measured moment for each case, normalized to the moment for total flux
      expulsion in the volume of the sample.
      The relation between both kinds of quantities is given by
      ${M\over H}={\overline{\chi} \over {1-N}}={\chi \over {1+4\pi N\chi}}$.
      A relation of the same kind holds for $\chi_g$ and $\overline\chi_g$.
      With the use of Eq.~(\ref{eq-chi_j}), we finally obtain:
      \begin{equation}
      \chi_j = {{\overline{\chi} - \overline{\chi}_g} \over
      {(1-N\overline{\mu})\ {\overline{\mu}_g}}}
      \label{eq-chi_j2}
      \end{equation}
      where $\overline{\mu} = 1+4\pi \overline{\chi}\,$,
      $\,\overline\mu_g = 1+4\pi \overline\chi_g\,$.
      Similar to the case of the sample A, an approximate curve has been determined for $\overline{\chi}_g$
      by interpolating between the small $H_{dc}$ curves at high temperatures, and
      the curve at $H_{dc} = 20\,G$ at low temperatures. Then the values of
      $\chi_j$ have been derived from Eq.~(\ref{eq-chi_j2}) and plotted on
      Fig.~\ref{bkhijic}. The set of curves is similar to
      the set for sample A, except for the higher onset temperature of the
      intergrains currents.

\begin{figure}[htb]
\epsfxsize=\hsize \epsfbox[0 0 400 300]{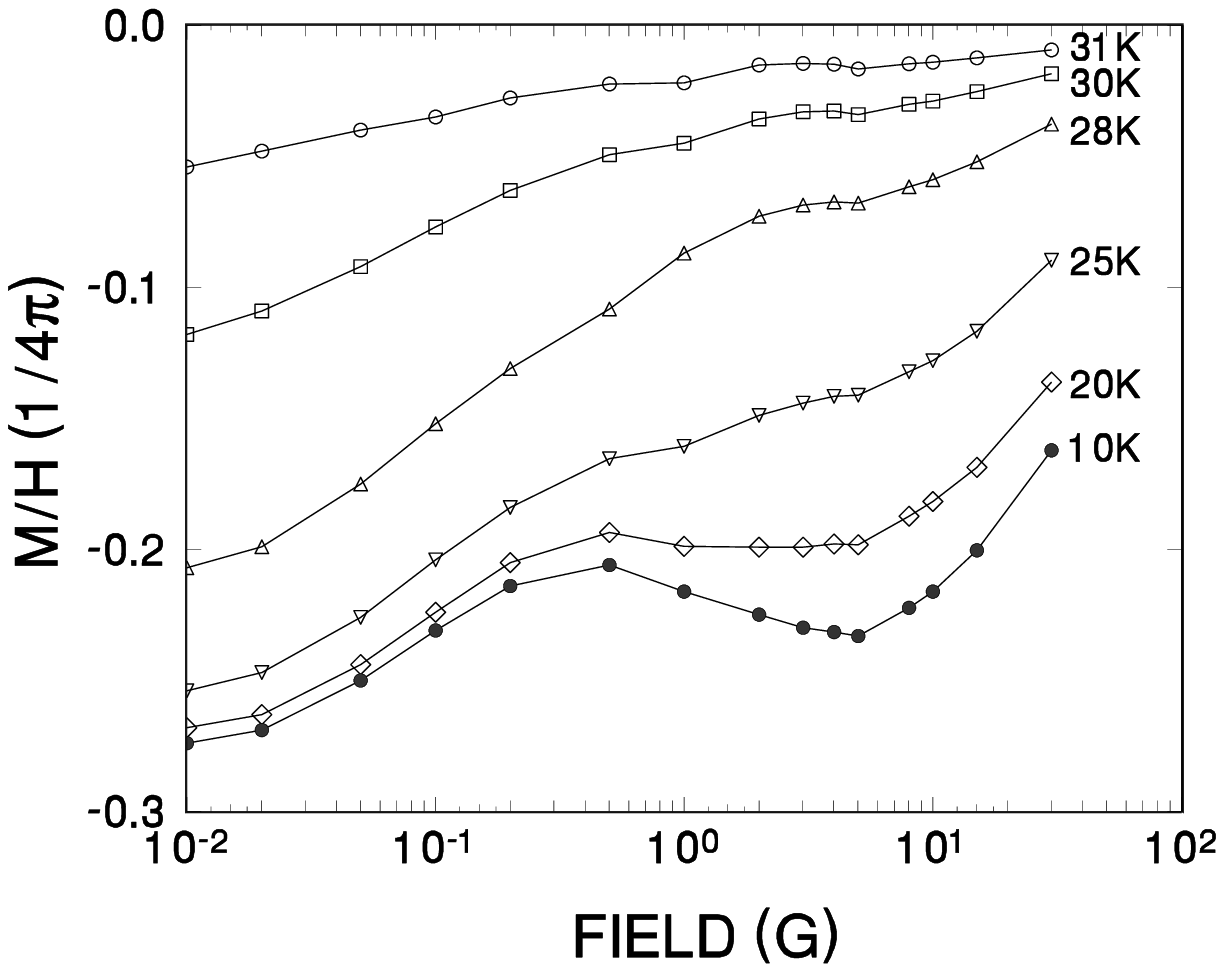}
\caption{F.C. (Meissner) susceptibility of sample B normalized to 
the moment for total flux expulsion, as a function of field.}
\label{bmeissvsh}
\end{figure}

      Field Cooled magnetization data, normalized to the value of the moment for full
      flux expulsion, are reported in Fig.~\ref{bmeissvsh} as a function of
      fields up to $30\,G\,$. 
      
 At the
      lowest field and temperature, the F.C. magnetization does not exceed $28\%$
      of its value for full flux expulsion. Furthermore, at low temperatures
      the curves versus field present a second maximum at about $5\,G\,$.
      We expect that this complicated behavior is due to the back
      shielding effect of the intergrain currents, as discussed
      for sample A. To take them into account,
      a relation similar to Eq.~(\ref{eq-chi_fc}) ( but with the demagnetizing 
       effect taken into account) should be derived. The internal field is given as usual by
      $H_i = H - 4\pi NM\,$,
      and the value of the local field seen by the
      currents is 
      $H_l = \mu_g \,H_i\,$.
      Thus, under a temperature variation $dT\,$,
      $${dH_l \over dT} = {d \mu_g \over dT}\,(H-4\pi NM)\
      -\ 4\pi N\ \mu_g\,{dM \over dT}\,.$$
      With  
      ${dB / dT} = \mu_j\, {dH_l / dT}\,$, 
      and using the relation
      $$M = {{\overline{\chi}_{FC} H}\over{1-N}} =
      {1 \over 4\pi}\ \int^T_{T_c}{d(B-H_i) \over dT}\,dT$$
      one obtains after integration: 
      \begin{equation}
      \overline{\chi}_{FC} = {{1-N} \over 4\pi N}\left(1-\exp
      \left(- 4\pi NI\right)\right) ;\qquad 
      I = \int^T_{T_c}{\mu_j \over
      {1-N\ (1-\mu_g\ \mu_j)}}\>
      {{d \chi_g\over dT}}\,dT\,.
      \label{eq-chi_fc2}
      \end{equation}
Here $\mu_j = 1+4\pi \chi_j$, with $\chi_j$ reported on Fig.~\ref{bkhijic},
whereas the value of $\mu_g$ was obtained using the relation 
      $\mu_g =(1-N) \overline{\mu}_g/ (1-N\overline{\mu}_g)$
      from the value of $\overline{\chi}_g$ as derived above.

      The values of $\overline{\chi}_{FC}$ for $H\rightarrow 0$ have been
      calculated using the values of $\overline{\chi}_g$ as determined above, and the values of
      $\chi_j$ at $H_{dc}=0$. The calculated value of $\overline{\chi}_{FC}$
was found to be about $-0.35$ at $T=10 K$, whereas its measured value
was about $-0.28$.
      The discrepancy between measured and calculated values
      is larger here than in corresponding results for sample A. We believe
      that the origin of this discrepancy is the same as in the 
      case of sample A, i.e. it stems from the nonlinear response effect. This effect
      is numerically larger in sample B since here
      the onset of Josephson
      currents occurs in a range of temperature where $\chi_g$ still varies
      strongly, contrary to the case of sample A.

 The above analysis
shows (irrespectively to the above-mentioned discrepancy)  that the back shielding
effect leads to a strong
      reduction of the field cooled susceptibility as compared with the
      susceptibility of the grains alone. It is then easy to understand the
      complex behavior of $\overline{\chi}_{FC}$ as a funtion of field:
      at $10\,K$ for instance,
      the onset of back shielding occurs at about $20\,G\,$, and its amplitude
      increases with decreasing field due to the increase of
      $\overline{\chi}_j\,$. Starting from the two-step behavior of $\chi_g$ 
      expected from the theory of Wohllebeen et al. \cite{Wohll} (and seen in the
      data of sample A, where back shielding is less important), back shielding
      results on the double maximum shape of the measured curves.
\section{Detailed Study of the Josephson Network Response.}
\subsection{Determination of the global critical current.}

In this subsection we will present the procedure we used to extract the value
of the macroscopic critical current in our sample B. This procedure is
not quite trivial since we are interested in the dependence of the critical
current on the background d.c. field in the sample, so we need to
analyse the magnetization curves which depend both on the cooling field
$H_{d.c.}$ and on the field variation $\delta H$.

\begin{figure}[htb]
\epsfxsize=\hsize \epsfbox[0 0 400 300]{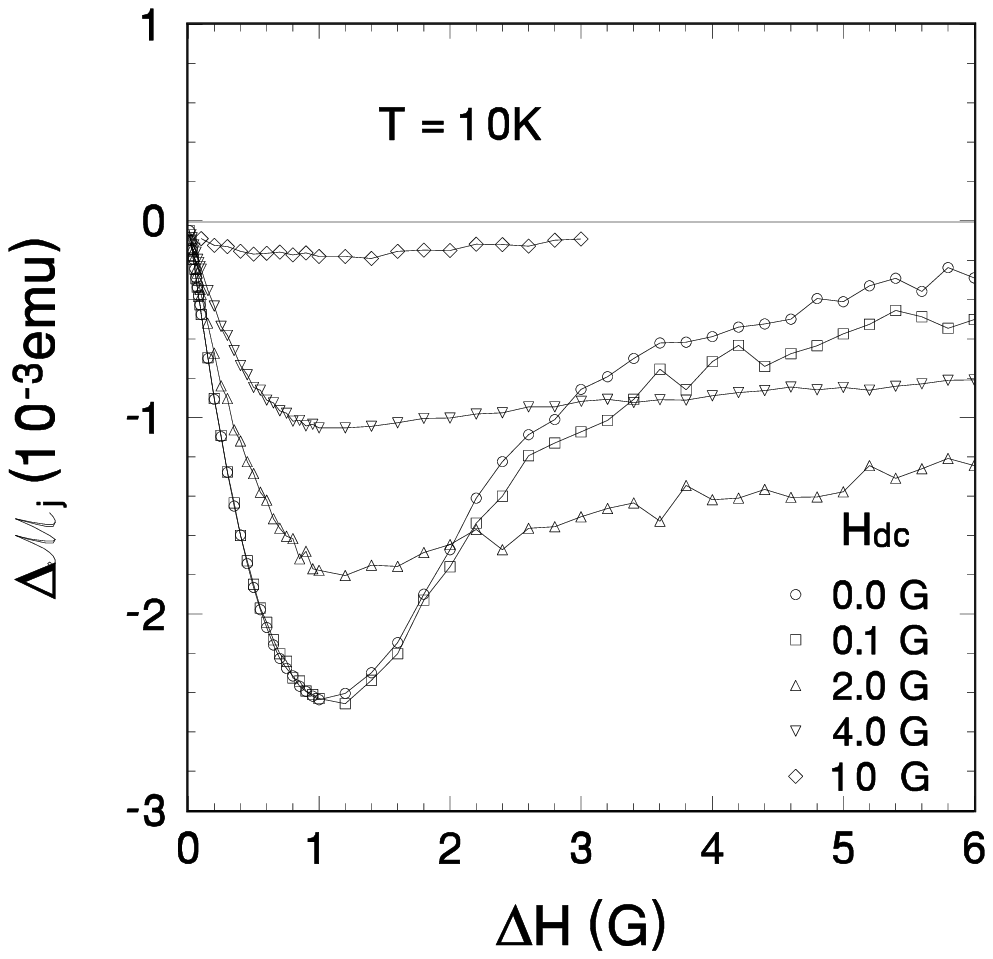}
\caption{Shielding moment of the Josephson currents after cooling the
sample at $10\,K$ in a d.c. field in the range $0-10\,G\,$.}
\label{minij_10}
\end{figure}

 Magnetization has been recorded at 10 and $20\,K$ as a function
of increasing $\Delta H$ with the smallest possible field steps
($\delta H =10\,mG$), and starting from several F.C. states.
From the $\Delta{\cal M}$ data, it is possible to derive the value of the
current response $\Delta{\cal M}_j$ as a function of $\Delta H$. For this,
we use Eq.~(\ref{eq-chi_j2}) which can be written as :
\begin{equation}
\Delta{\cal M}_j = {{\Delta{\cal M} -\Delta{\cal M}_g}\over
{(1-N\overline\mu )\,\overline{\mu}_g}}\,.
\label{dMj}
\end{equation}
where $\Delta {\cal M}_g$ is the magnetization of the grains alone;  
$\overline \mu$ and $\overline \mu_g$ are defined in Section II-B.
The value of the grains system response is approximately derived in the same section:
$\Delta {\cal M}_g \approx 3.2\,10^{-3}\times H\,cm^3-G$ at $10\,K$ and
$\Delta {\cal M}_g \approx 2.9\,10^{-3}\times H\,cm^3-G$ at $20\,K$.
Calculated values of $\Delta{\cal M}_j$ at $10\,K$ are plotted
in Fig.~\ref{minij_10}. Note the analogy of these results to 
the magnetization curves of classical type II superconductors with strong
pinning (the difference is that here $\Delta H$ plays the role of
$H$).

\begin{figure}[htb]
\epsfxsize=\hsize \epsfbox[0 0 400 300]{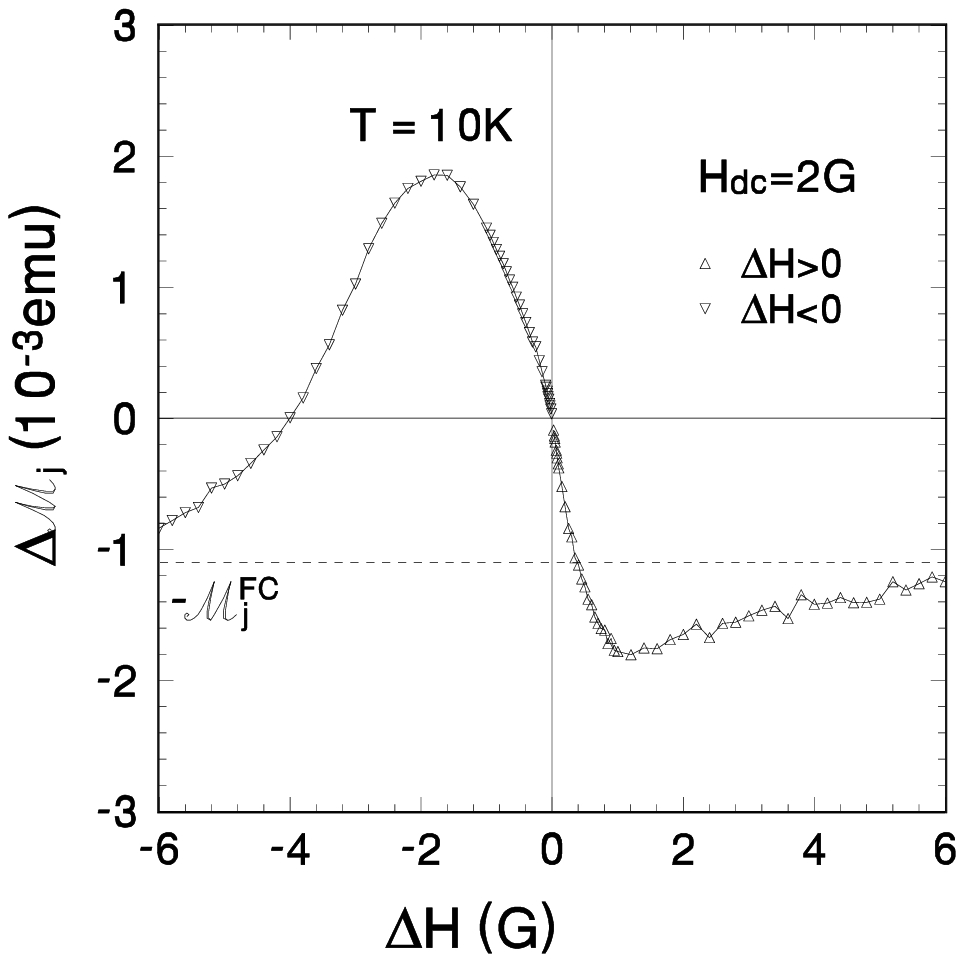}
\caption{Shielding moment of the Josephson currents after cooling the
sample at $10\,K$ in a d.c. field $H_{dc} = 2\,G\,$. Data are for positive
and negative field steps.}
\label{mpm2l}
\end{figure}

After cooling the sample at zero d.c. field, its response is obviously
symmetric with respect to $\Delta H$. When it is cooled in a finite
d.c. field, it is not the case anymore, as was explained 
 in the previous section.
The magnetic moment of the sample just after cooling is
${\cal M}_{FC} = {\cal M}_g + {\cal M}^{FC}_j$ where ${\cal M}^{FC}_j$ is
the positive moment due to the back shielding by the Josephson currents
which have been developed during the cooling process
(see Eq.~(\ref{eq-chi_j} and~\ref{eq-chi_j2})). So, the total moment
produced by the intergrain
currents after increasing the field by $\Delta H$ is
${\cal M}_j = {\cal M}^{FC}_j +\Delta{\cal M}_j$. It
is this moment which vanishes when $J_c \rightarrow 0$ (at large enough
$\Delta H$), and thus $\Delta{\cal M}_j$ approaches $-{\cal M}^{FC}_j$.
In Fig.~\ref{mpm2l} we show the data recorded at $T=10\,K$ and $H_{dc} =2\,G$.
Curves recorded at positive and negative $\Delta H$ both converge to the
value corresponding to $-{\cal M}^{FC}_j$: at $10\,K\,$, 
$-{\cal M}^{FC}_j$ is about $1.1\,10^{-3} emu$.

When $\Delta H > H_{dc}\,$, it is natural to expect that the response of the
Josephson network does not depend on the initial state. A simple
illustration can be given by analogy with Bean-like pinning in type II
superconductors \cite{Bean}.
At large $\Delta H\,$, when the induction profile has penetrated up to the
center of the sample, the magnetization no longer depends on  $\Delta H$
but only on $J_c$. If, as it is the case in real materials, $J_c$ varies
with the induction in the sample, the magnetization depends on the total
$H\,$, whatever the value of $H_{dc}$ in which the sample was cooled.
Actually, when plotted as a function of the total field
$H_{dc}+\Delta H\,$, the curves giving the total moment of network currents
${\cal M}^{FC}_j +\Delta{\cal M}_j$ merge in their "large" field part (i.e. 
above their maximum). The values have been calculated, with
$-{\cal M}^{FC}_j = 1.1\,10^{-3} emu$ and $0.7\,10^{-3} emu$ for $H_{dc}=2\,G$
and $4\,G$ respectively. In order to obtain an optimal overlap between the
curves, the following values have been used for $\Delta {\cal M}_g\,$:
$3.25\,10^{-3}\times H\,cm^3-G$ at $H_{dc}=0\,G\,$,
$3.22\,10^{-3}\times H\,cm^3-G$ at $H_{dc}=2\,G$ and $4\,G$. Indeed, the
calculated values for $\Delta{\cal M}_j$ at large $\Delta H$ are extremely
sensitive to those for $\Delta {\cal M}_g\,$. This allows us to refine the
determination of $\Delta {\cal M}_g\,$. Note that the values quoted above
do not differ by more than $1\%$, which is compatible with experimental
accuracy and the possible variations of grains response with $H_{dc}\,$.

\begin{figure}[htb]
\epsfxsize=\hsize \epsfbox[0 0 400 300]{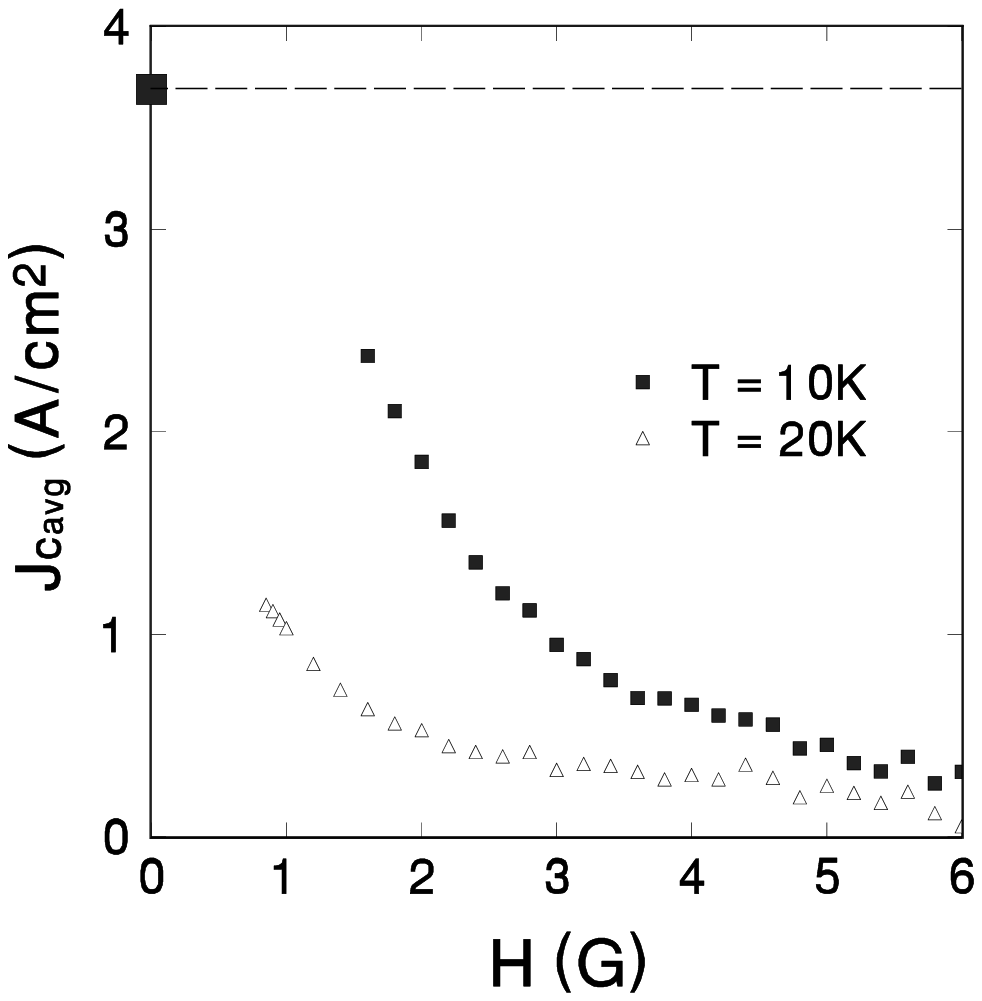}
\caption{Calculated values of the averaged critical current ${\tilde J}_c$
as a  function of total field for strong field penetration. The big square 
corresponds to the initial $J_c$ as determined in III-B}
\label{jcavg}
\end{figure}

Finally, from the knowledge of the true Josephson shielding response, in 
``large'' fields we can now derive a rough evaluation of the critical current.
Namely, above the maximum of $\Delta {\cal M}_j\,$, we calculate the value
${\tilde J}_c$ of the average critical current which would give the value of 
the measured moment by use of the Bean formula \cite{Bean} in a cylindrical geometry.
For strong penetration, the magnetization is given in e.m.u. by
$M={\tilde J}_c R/3$.
With $R=0.15\,cm$ and the values of the moment measured
at $10\,K$ and $20\,K$ with $H_{dc}=0\,G\,$, we obtain the data displayed
in Fig.~\ref{jcavg}. Note that the data are limited to fields such that
$H\approx H^* = 4\pi {\tilde J}_c R$ below which the above approximate evaluation
is no longer relevant.

\subsection{Low Field d.c. Response.}
\begin{figure}
\epsfxsize=\hsize \epsfbox[0 0 400 300]{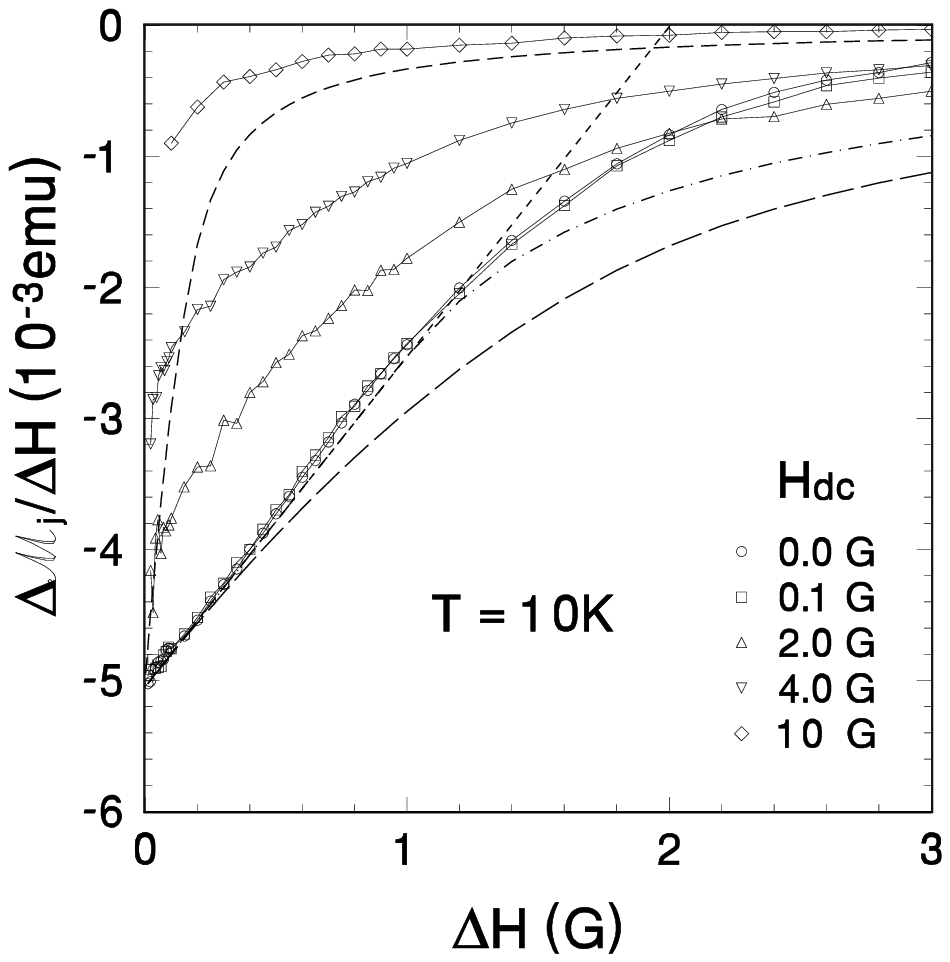}
\caption{Josephson currents susceptibility at $10\,K\,$ vs. applied variation 
$\Delta H$ of field, after cooling in d.c. field $H_{dc}\,$. The meaning of 
dashed and dot dashed lines is explained in the text.}
\label{xinis_10}
\end{figure}

\begin{figure}
\epsfxsize=\hsize \epsfbox[0 0 400 300]{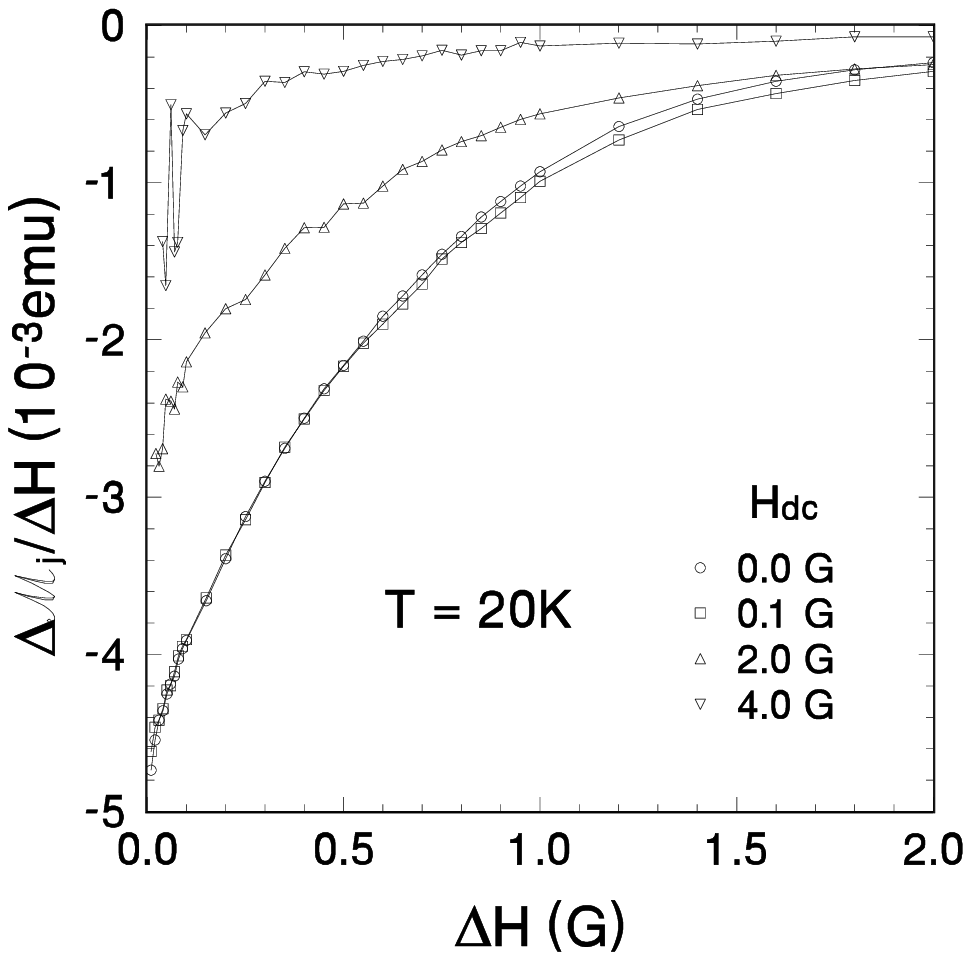}
\caption{Josephson currents susceptibility at $20\,K\,$ vs. applied variation 
$\Delta H$ of field, after cooling in d.c. field $H_{dc}\,$.}
\label{xinis_20}
\end{figure}

We can now concentrate on the behavior of the Josephson currents moment at
small $\Delta H\,$. For this discussion, the currents susceptibility
$\Delta{\cal M}_j /\Delta H$ is plotted versus $\Delta H$ at $10\,K$ and
$20\,K$ in Fig.~\ref{xinis_10} and~\ref{xinis_20}, respectively.


At $10\,K\,$, after zero field cooling or cooling in a small field
$H_{dc}=0.1\,G\,$, the response varies linearly with $\Delta H$ for
small values of $\Delta H$ up to about $0.5\,G\,$. This linear slope
of $\Delta{\cal M}_j /\Delta H$ 
is considered as the result
of classical Bean-like pinning with critical current density
$J_c = H^*/4\pi R\,$, where $1/4\pi H^*$ is the initial slope of the curve
\cite{Bean}. This initial slope is reported on
the figure as the short dashed line which corresponds to $H^* = 2\,G$,
leading to $J_c \approx 3.7 A/cm^2$.

At larger $\Delta H\,$, the behaviour of currents susceptibility
$\Delta{\cal M}_j /\Delta H$ deviates from linear, which is the result
 of both the magnetic-field dependence of the critical current $J_c\,$,
 an intrinsic effect, and
of the increasing degree of flux penetration into the sample
, a purely size-dependent effect. Usually one uses the Bean model
(generally with some $B$-dependent critical current) in an appropriate geometry
in order to deconvolute these two effects. However, one should keep in mind
that
the Bean model is a severe simplification of the problem of constant
pinning force, corresponding to the limit $\lambda \to 0\,$ (i.e. the London
penetration depth is supposed to be negligible with respect to the Bean
penetration length). For the simplest sample shapes (thin slab or
cylinder) it means that the condition $\lambda \ll R$ should be fulfilled,
which is usually the case. However the situation is more complicated for
samples of square cross-section (like our one), where the effect of
corners may become important even at $\lambda \ll R$. For such a geometry, 
the use of Bean model leads to exactly the same relation between critical
current, external field and measured magnetization as for the cylindrical
ones, whereas one expects some difference if finite-$\lambda$
corrections are taken into account. At the present stage, we are not 
able to evaluate these corrections and therefore the values of the 
magnetization 
corresponding to our experimental geometry with non negligible $\lambda\,$.  Nevertheless, we  
expect that it lies between the curves for two extreme limits. The upper one 
corresponds to the $\lambda \to 0$ limit, where the magnetization is given 
simply by the Bean's formula for 
the cylinder: $4\pi M/H= -1 + H/H^* - H^2/3{H^*}^2\,$ for $H<H^*$ and 
$4\pi M/H= -H^*/3H$ for $H>H^* \,$. A lower limit (thought rather artificial) 
consists of the "double slab" case 
in which the variation of magnetization is counted twice (once for each 
pair of edges): $4\pi M/H = -1 + H/H^*$ for $H<H^*/2$ and 
$4\pi M/H = -H^* /4H\,$ for $H > H^*/2$. 
Both curves are plotted in the Fig.~\ref{xinis_10} (dot dashed
and long dashed curves respectively) for $J_c = 3.7 A/cm^2$ and
$\Delta{\cal M}_j /\Delta H = -5.05\,10^{-3} cm^3$ at $\Delta H \to 0$. 

Let us now discuss the data starting from those obtained for low d.c. fields,
$H_{d.c.} = 0, 0.1 \,G$. One can
see that, after the initial linear part, the absolute value of the measured
susceptibility
is always smaller than the calculated ones. This
corresponds to the decrease of $J_c$ with increasing induction, as it is
classically expected in granular materials, due to the suppression of
intergrain critical currents by magnetic field penetration
into the Josephson junctions\cite{Vela}.
This ``classical'' behavior for granular superconductors is usually analyzed
by considering the volume-averaged Josephson medium as a kind of type II
superconductor in the dirty limit, provided its macroscopic
penetration depth $\lambda_J$ is large as compared with the grains size
\cite{sonin,Clem}.

At $H_{dc} \ge 2\,G$  the behavior of $\Delta{\cal M}_j /\Delta H$ is
quite different: there is no initial linear slope, but a monotonic
curvature is present down to the smallest $\Delta H\,$.
It is no longer possible to adjust a Bean like curve to the data. For instance, 
the Bean curve plotted on the lowest $\Delta H$ data for $H_{dc}=2\,G$ is 
reported on the Fig.~\ref{xinis_10} as a dashed line. It corresponds to a 
very small critical current of order $0.2\,A/cm^2\,$,
 and it is evident that 
the effective screening current becomes much larger with increasing 
$\Delta H\,$. 
Here, contrary to the case of $H_{dc} =0G\,$, the absolute value
of the measured susceptibility is always {\it larger} than the calculated
one for a constant shielding current corresponding to the limit 
$\Delta H \to 0\,$. This means that, whereas at
$H_{dc}=0$ the effective screening current density stays constant and then
slowly decreases with increasing $\Delta H\,$  (which corresponds to classical
Josephson pinning), at $H_{dc} \ge 2\,G$  it {\it increases} with
$\Delta H\,$ {\it sublinearly} (since a linear increase would correspond
to a susceptibility independent of $\Delta H$). Such behaviour
is quite unusual within the commonly accepted
picture of screening in superconductors; indeed, we know that, for vanishing 
field excitations, the screening
current may be either i) {\it linear} in $\Delta H$ and {\it reversible}, as in the London (or Campbell \cite{Campbell}) shielding regime,
or ii)  {\it constant} (equal
to the initial critical current $J_c$) and {\it irreversible} as in the
case of the Bean-type critical state (or of any other known critical model, 
e.g Kim model \cite{kim}, exponential model \cite{expo}, etc).

\begin{figure}[htb]
\epsfxsize=\hsize \epsfbox[0 0 400 300]{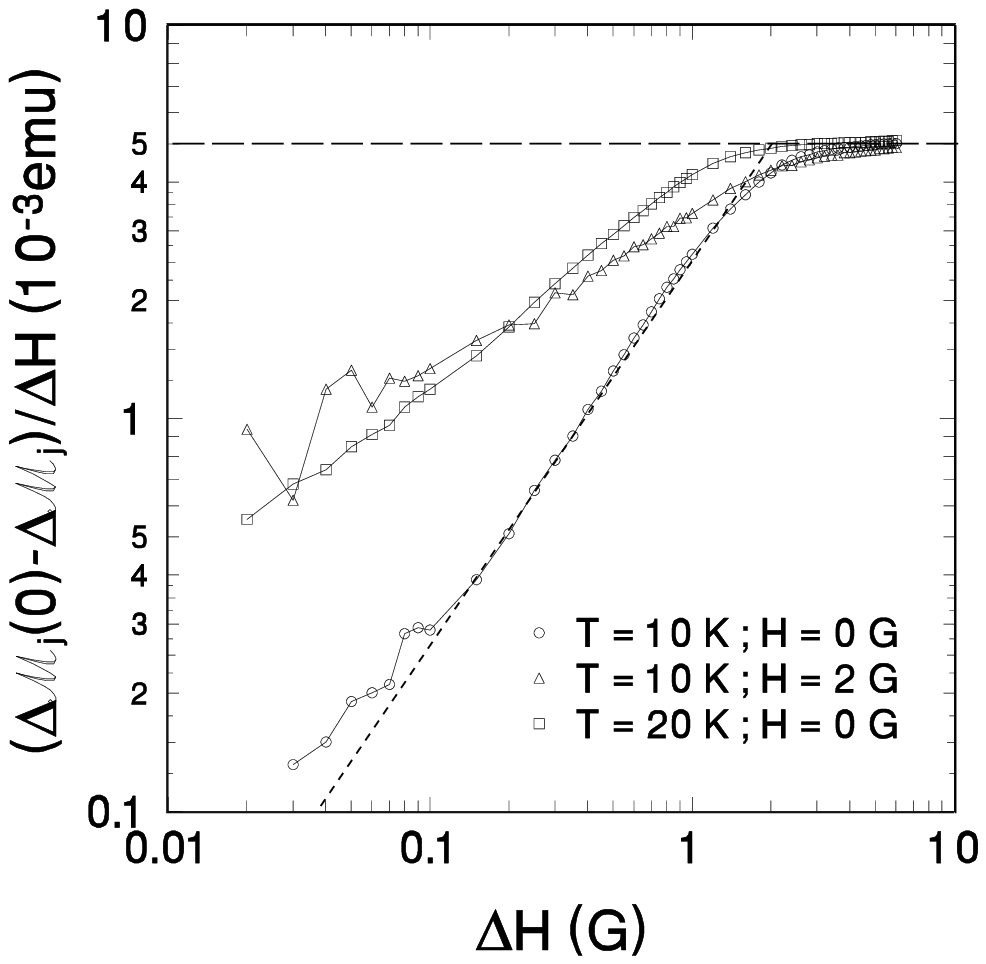}
\caption{Difference between the measured susceptibility and its
value for perfect shielding for selected data at $10\,K$ and $20\,K\,$.
The short dashed line represents a logarithmic slope 1 expected for a
Bean critical state.}
\label{dx2_1020}
\end{figure}

The above anomalous screening behaviour is even more pronounced
at $20\,K$ where, even after zero field cooling, no initial linear
slope of $\Delta{\cal M}_j /\Delta H$ can be seen in the data.
All curves show the
same anomalous behavior as the data at $10\,K$ in fields from $2\,G\,$.
This specific behavior is
emphasized by plotting the difference between the measured susceptibility
$\Delta{\cal M}_j /\Delta H$ and
its value for total flux expulsion $\Delta{\cal M}_j(0) /\Delta H$,
versus $\Delta H$ on a Log--Log scale. In such a plot, at least in the
regime of weak penetration, i.e. where
$\Delta{\cal M}_j /\Delta H$ is larger than
$0.8\,\Delta{\cal M}_j(0) /\Delta H\,$, sublinear variation of the shielding
current density results in a logarithmic slope smaller than 1 for the
curves of $\Delta{\cal M}_j /\Delta H\,$ (for $\Delta{\cal M}_j /\Delta H$
smaller than $0.8\,\Delta{\cal M}_j(0) /\Delta H\,$, we are in a regime of
strong penetration where it is no longer possible to relate simply the
variations of the moment response to those of the shielding current). In
Fig.~\ref{dx2_1020}, we have reported the three curves for which data are found in the range above $0.8\,\Delta{\cal M}_j(0) /\Delta H\,$, i.e. at
$T=10\,K\,,H_{dc} =0\,G$ and $2\,G\,$, and $T=20\,K\, H_{dc} =0\,G\,$.

At $10\,K$ and $H_{dc} = 0\,G\,$, the logarithmic slope is about 1
as expected, although at the smallest fields the curve crosses over
to a smaller logarithmic slope closer to  $0.5$.
At $20\,K$ and $H_{dc} = 0\,G\,$ the logarithmic slope is about $0.4$ 
at the lowest $\Delta H\,$. Approximately the same value of the slope characterizes the data obtained
at $10\,K$ and $H_{dc} = 2\,G\,$, although the dispersion of data points
at lowest $\Delta H\,$ makes its accurate determination difficult.

 The above anomalous behaviour makes it tempting to try a simple
{\it Ansatz} for the behavior of the response current density of the system
versus induction variations. Let us suppose that
$J \propto {\Delta B}^{\alpha}$ with $\alpha$ between $0$ and $1$.
The case with $\alpha = 1$ corresponds simply to classical screening with
penetration length $\lambda$ (since $J \propto \Delta B$); the case with
$\alpha = 0$ corresponds to constant $J$, i.e. the classical Bean case.
Anomalous response arises for non integer $\alpha\,$.
For very small excitation $\Delta H\,$,
the length of induction penetration is small as compared with the size of the
sample and we need to consider the effect of the excitation
in the lowest order in $\Delta B$ only. For the purpose of illustration we consider
the simplest slab geometry.
Then the induction profile is determined by the Maxwell equation
\begin{equation}
{dB\over dx} = -{4\pi J_1}\left(\frac{\Delta B}{\Delta B_1}\right)^{\alpha}
\label{dBdx}
\end{equation}
where $x$ is the coordinate perpendicular to the edge of the sample.
For an external field $\Delta H\,$, the induction in the sample
is given by
\begin{equation}
\Delta B(x) = \left( {(1-\alpha)4\pi J_1\over {{\Delta B_1}^{\alpha}}}
\,(x_H -x) \right)^{1/(1- \alpha)}
\label{dBx}
\end{equation}
where $x_H$ is the coordinate of penetration and $J_1$ and $\Delta B_1$ are
normalizing factors; for $x = 0\,$, $\Delta B=\Delta H$, i.e.
$x_H = {\Delta B_1}^{\alpha}/ (4\pi J_1 (1-\alpha))\cdot
\Delta H^{1- \alpha}\,$.
Then, integrating the field profile (\ref{dBx}) over $x$, we get
\begin{equation}
\frac{4\pi\Delta\overline M + \Delta H}{\Delta H}\propto {\Delta H}^{1-\alpha},
\label{dMdH}
\end{equation}
where $\Delta\overline M = \Delta{\cal M}/V$ is the mean magnetization
variation due to the field variation $\Delta H$.

If we now compare the result (\ref{dMdH}) with the data shown in
Fig.~\ref{dx2_1020}, we find values of $\alpha$ in the range $0.4 \div 0.5$
at both $10\,K$ and $20\,K$.

Thus a simple choice for the relation between the screening current
$J_c$ and the induction variation $\Delta B$
allows us to imitate the experimental results for the simplest protocol
of a weak monotonic $\Delta H$ variation on top of a homogeneous state of the
network.
Nevertheless, it is evident that $\Delta B$ has no clear meaning if the
variation of $H$ is non-monotonic or if the initial state is
obtained by non-zero field cooling. Indeed, in the later case, induction in the
sample varies during cooling due to the variation of $\mu_g$ with $T\,$,
giving the response $\Delta {\cal M}_j^{FC}$ as seen before.
Furthermore, we will see below that the response
 is irreversible even for extremely low exitation fields.

\subsection{Irreversibility: Very Low Field, low Frequency a.c. Response.}
Problems of sensitivity limit the range of small excitations
which can be used in d.c. experiments. The preceding results clearly show
the sublinear nature of the low field response, but they do not allow its
precise determination.
In order to extend by several orders of magnitude the range of our lower excitations investigation, we have been led to perform a.c. susceptibility
measurements. The use of a.c. response measurements is always questionnable
when equilibrium (or quasi-equilibrium) properties are under investigation,
since the results can be affected by the time-dependent
part of the response function. It has been shown that the latter is the
response of a very good conductor with complex conductivity \cite{Leylek2,%
Leylek1}. Hence, it is necessary to work at low frequency, in a range where
the susceptibility is roughly frequency independent.

\begin{figure}[htb]
\epsfxsize=\hsize \epsfbox[0 0 400 300]{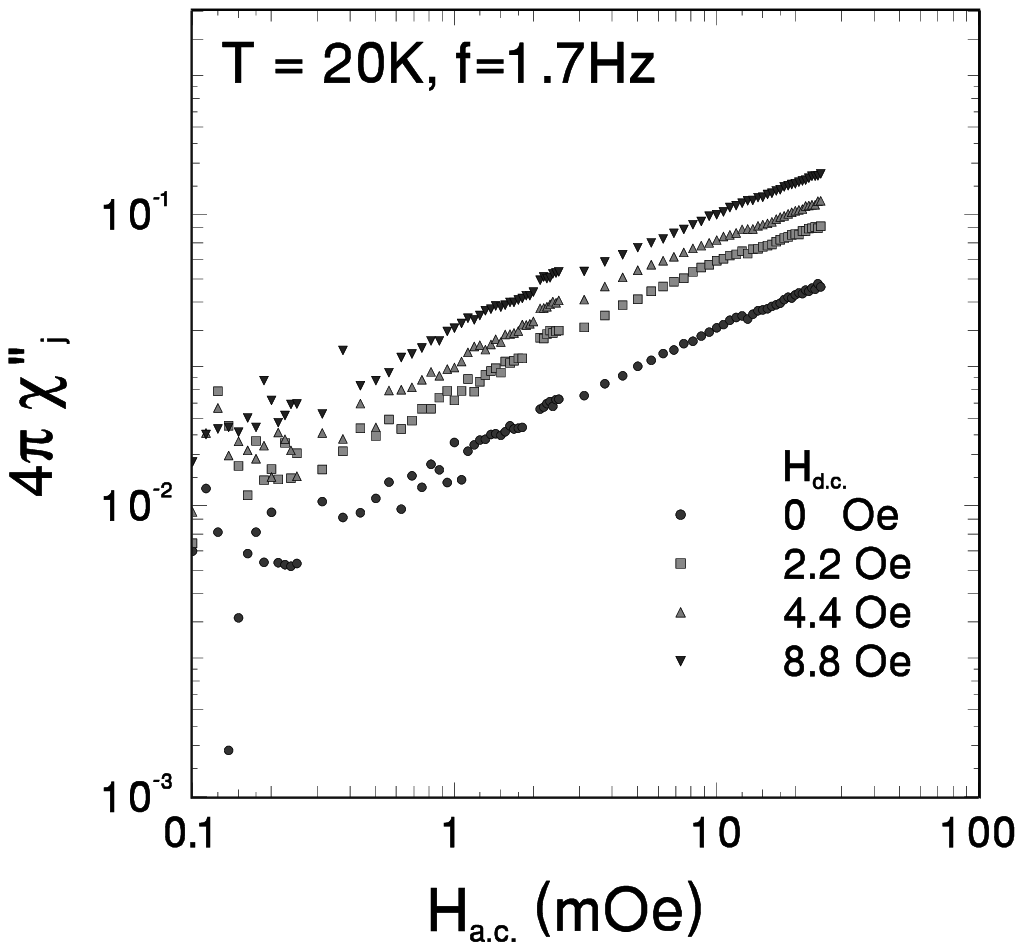}
\caption{Out-of-phase susceptibility at $1.7\,Hz$ as a function of a.c.
field amplitude.}
\label{xsjvslh}
\end{figure}

We present here preliminary results obtained on a long cylinder obtained 
by stacking several of the original sample B 
cylinders. Measurements were done at $20\,K$, at a working frequency of $1.7\,Hz$ in the equipment used for noise experiments 
\cite{Leylek1}. The sample was simply shifted into the upper half of 
the third order gradiometer. At this temperature and
frequency, we have verified that the in-phase susceptibility is almost
frequency independent, which ensures that the results are mainly dependent
on the (quasi) static part of the response. The susceptibility was recorded using classical method of SQUID magnetometry. We used a.c. excitation fields
in the range $3\,10^{-2}$--$30\,mG$ and the sample was cooled in d.c.
fields from $0$ to $8.8\,G$. From the data, the values of the Josephson
network susceptibility was extacted using the method developed in Section II,
with the susceptibilities in Eq.~(\ref{eq-chi_j2}) being complex 
quantities. The susceptibility measured at $4.2\,K$ at the lowest a.c. 
amplitude was taken as the level for perfect diamagnetism. Fig.~\ref{xsjvslh} displays a log-log plot of the out-of-phase 
susceptibility $\chi''_j$ versus the amplitude of the a.c. field, and for 
several values of the F.C. static field.  The response is irreversible 
down to the lowest a.c amplitudes, and the irreversibility increases with 
the superimposed d.c. field. 
\begin{figure}[htb]
\epsfxsize=\hsize \epsfbox[0 0 400 300]{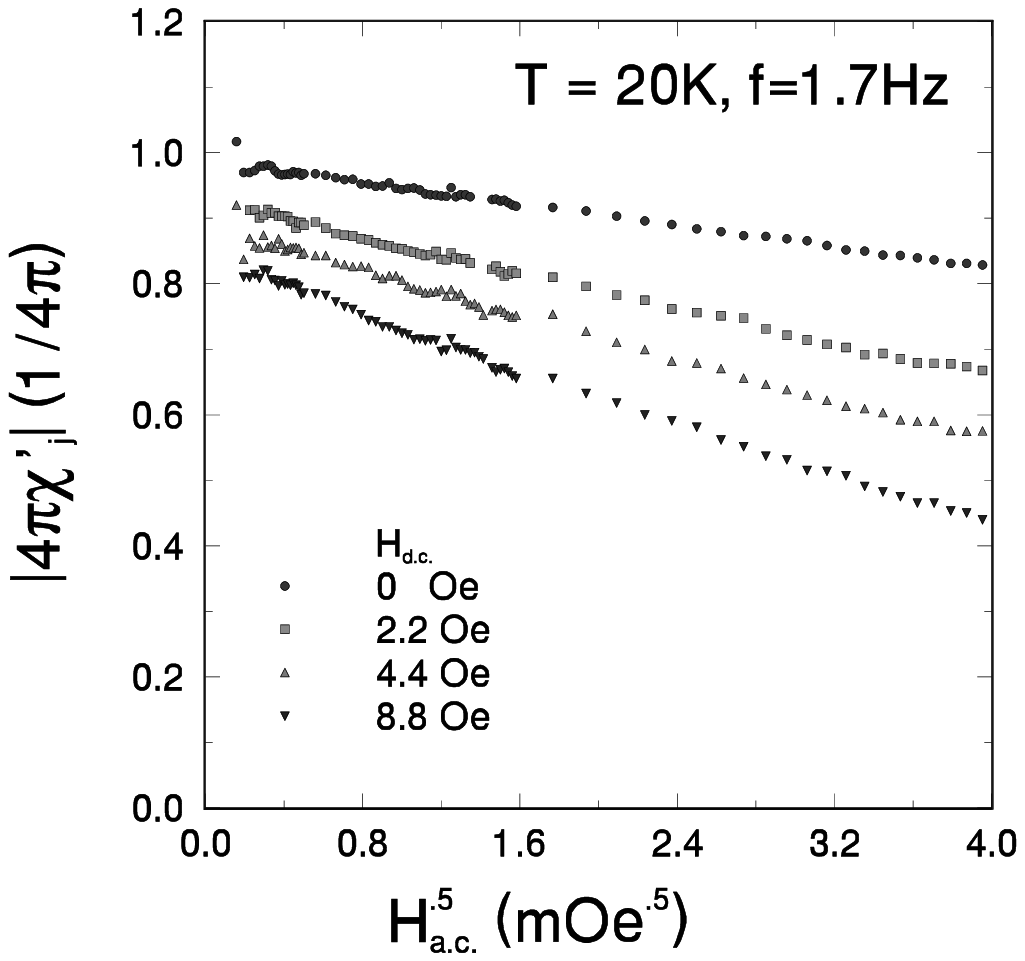}
\caption{In-phase susceptibility at $1.7\,Hz$ as a function of the power of 
a.c. field amplitude $H_{ac}^{0.5}\,$.}
\label{xpjvsla}
\end{figure}
All curves follow a power law 
, with the same exponent close to $0.5\,$. Going towards the smallest 
a.c. excitations, they show some downward bend which could be related 
with the approach to a linear regime (with $\chi''_j =0$) below $0.1\,mG\,$, 
although the dispersion of the data is too high to conclude.
The in-phase susceptibility $\chi'_j$ is plotted as a function of 
$H_{ac}^{0.5}$ in Fig.~\ref{xpjvsla}. Here as well, the anomalous nature of the 
response is clearly seen. 
$4\pi\chi'_j$ behaves like 
$(-1+\delta +\gamma H_{ac}^{0.5})$ where both the constant $\delta$ and the 
slope $\gamma$ increase with increasing superimposed static field $H_{dc}\,$.
The dependence of the harmonic susceptibility on the a.c. field amplitude
is a genuine proof of the existence of static irreversibility in the
response. This is not astonishing by itself, but these results stress
the anomalous aspect of this irreversibility. For instance, in the classical
Bean case with a weak penetration, it is known that $1+4\pi\chi'_j$ and
$\chi''_j$ are proportional to $H_{ac}\,$ whereas Fig.~\ref{xsjvslh} and 
~\ref{xpjvsla} clearly show the proportionality to $H_{ac}^{0.5}$. 
A further evidence is provided by plotting $\chi''_j$ versus
$1+4\pi\chi'_j$ as displayed in figure~\ref{dxpvsxs}. It can be shown that
if the a.c. response is driven only by static irreversibility, both are
proportional. In the Bean case, the coefficient of proportionality is
$4/3\pi\,$. In the figure~\ref{dxpvsxs}, the part of data which lie in the
range of $20\%$ variation of $\chi'_j$ (where the relations for slab geometry
are approximately valid) show that $\chi''_j$ is indeed
proportional to $1+4\pi\chi'_j$, but with a bit smaller coefficient 
$\approx 0.28 \pm 0.03\,$. 
\begin{figure}
\epsfxsize=\hsize \epsfbox[0 0 400 300]{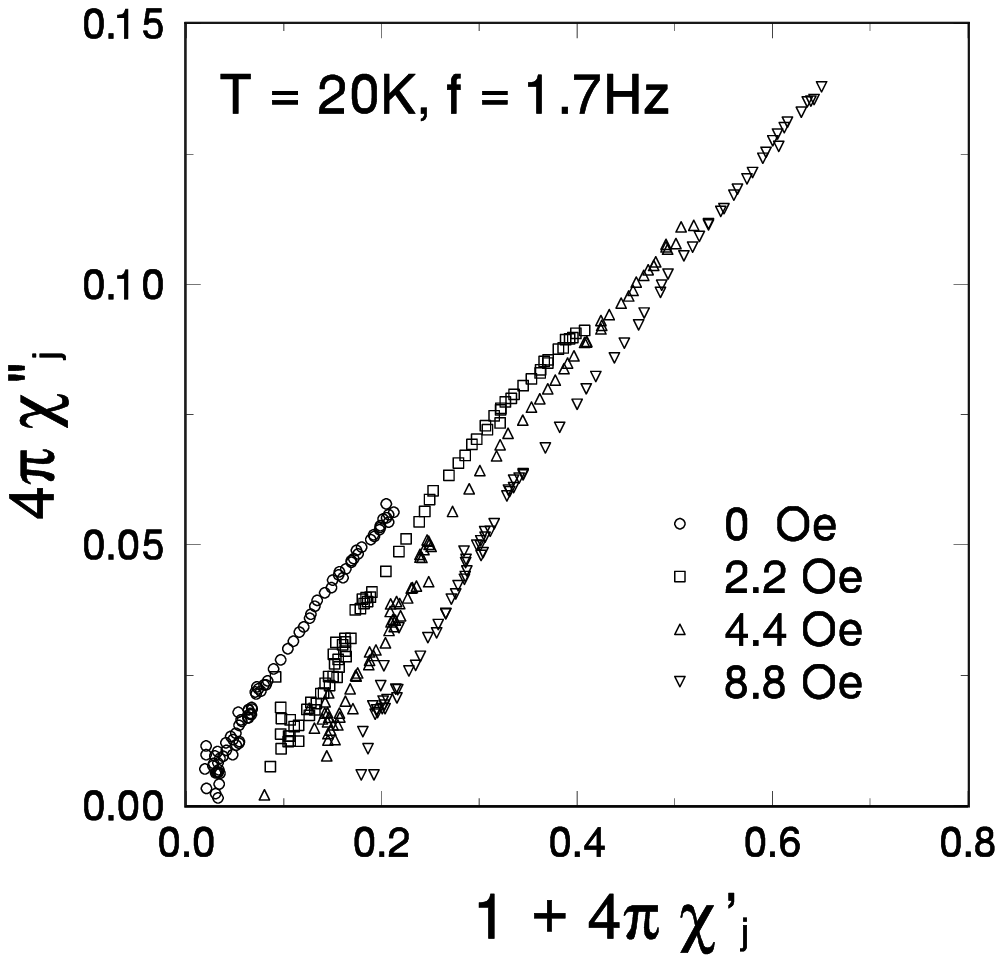}
\caption{Plot of $4\pi\chi''_j$ as a function of $1+4\pi\chi'_j$ for
the different values of $H_{dc}\,$.}
\label{dxpvsxs}
\end{figure}

In order to understand the meaning of the above results, we
generalize the crude {\it ad hoc} model of Section III-B
to the irreversible case.
 In order to do it, we  generalize the protocol of the Bean
model. Namely, in the Bean model, the current is given by a step function of
 the
variation of induction, $J = J_c sgn(\Delta B)$ according to the 
sign of $\Delta B$, as long as 
the induction variation is monotonous. 
If the sign of variation of $B$ is reversed, 
$J$ also changes sign, which can be written in terms of the {\it variation}
of the current density (with respect to the initial current distribution obtained after monotonous variation of the field, $J_{init}\,$,
$\Delta J = -2J_{init} \Theta(-\Delta B_{new})\,$
 where
 $\Theta(x)=(1/2)(1+sgn(x))$ and $\Delta B_{new} = B-B_{init}\,$.
Such a representation (which is not needed in the analysis of
 the Bean model itself) will allow us to construct the necessary generalization
of the relation between current and variation of the field used in Eq.(~\ref{dBdx}). Actually our goal here is rather limited: we are going to
find a consistent description of the simplest hysteresis cycle which consists of the initial increase of $\Delta B$ from zero to $\Delta B_{init}$,
then reversing the sign of the field variation until the value of $\Delta B =
- \Delta B_{init}$ is reached, and then reversing $dB/dt$ once more and
finishing at  $\Delta B_{final} = \Delta B_{init}$. The description of this
cycle will be consistent if we find that the value of the current density
at the end-point, $J_{final}$, coincides with the one after the original increase
of the field
 $\Delta B_{init}$, $J_{init}\,$. This simply means that the
hysteresis loop is closed. It is easy to check that the above condition will
be fulfilled by the following choice of the $\Delta J(\Delta B_{new})$ 
dependence: 
\begin{equation}
\Delta J = -sgn(J_{init}) 2^{1-\alpha} J_1 \left({\Delta B_{new}}\over\Delta B_1\right)^{\alpha} \Theta(-\Delta B_{new})\,.
\label{JBnew}
\end{equation}
where $J_1$ and $\Delta B_1$ have the same meaning as in Eq.(\ref{dBdx}).
Actually the only difference between the Eq.~(\ref{JBnew}) and the original
used in the Eq.(\ref{dBdx}) is the coefficient $2^{1-\alpha}$.
The Bean model limit then corresponds to $\alpha \rightarrow 0$, so the above coefficient approaches 2 as it should.
Then instead of Eq.~(\ref{dBdx}) we obtain:
\begin{equation}
{d\Delta B }/ dx =\pm 2^{1-\alpha} {\cal A}\,{\Delta B}^{\alpha}
\label{dBdx2}
\end{equation}
where ${\cal A}=4\pi J_1 /\Delta B_1^{\alpha}\,$.
The induction profile, induced magnetization and harmonic response are calculated in the 
Appendix.
The main conclusions are that the fundamental
components $1+4\pi\chi'_j$ and $\chi''_j$ are both proportional to
${h_o}^{1-\alpha}\,$, and that their ratio $R= \chi''_j/(1+4\pi\chi'_j)$
decreases from $4/3\pi$ to $0$ when
$\alpha$ goes from 0 to 1. For $\alpha = 0.5\,$, we get
 (cf. Fig.~\ref{pentxsxp})
 $R\approx 0.25\,$, a
value which is in good agreement with the data presented on
Fig.~\ref{dxpvsxs}. Note that the degree of irreversibility
(measured by this ratio) is similar (although a bit lower)
 to the one of the Bean model.
It should be emphasized that the numerical coefficient in
 Eq.~(\ref{JBnew}) was ``fitted'' in order to obtain consistent
 (i.e. closed) hysteresis loop; one can expect that an analogous equation
describing current variation after some more complicated history of the field
variations will contain another (history-dependent) numerical coefficient
instead of $2^{1-\alpha}$. 

It can be seen from Figs.~\ref{xsjvslh} and~\ref{xpjvsla} that 
${\cal A}=4\pi J_1 /\Delta B_1^{\alpha}$ increases with increasing
 ambient d.c. field. It 
is natural to expect a decrease of $J_1$ with increasing $H_{dc}\,$. The increase of $\cal A$ with $H_{dc}$ means that 
$\Delta B_1^{\alpha}$ decreases more quickly than $J_1$ when $H_{dc}$ 
increases. 
The presence and behavior of the constant $\delta$ cannot 
be predicted on the basis of the above simple model. In fact, the latter neglects the possibility of elastic displacement of flux lines under the 
action of the external applied field. Such an effect would result in a  
response analogous to the Campbell response  due to 
the elastic displacement of vortices in their pinning potential in type II 
superconductors~\cite{Campbell}. Campbell's response is linear, so it would result in an a.c. field independent 
positive contribution to $\chi'$, whose amplitude would be inversely 
proportional to the strength of the  
restoring force of the pinning potential wells. It is natural to expect that 
the pinning force decreases with increasing ambient d.c. field in our granular 
system, due to the reduction of the junctions critical currents. Hence, such 
an effect would give a positive contribution to $\chi'$, which would increase 
with increasing d.c. field. This corresponds rather well to the behavior of the 
offset $\delta$ seen in the data.

\section{Comparison with an existing theory of gauge glass: frustration at
$H=0$}

In this Section we compare the experimental results described above with
the theoretical results available for the randomly frustrated Josephson networks.
We start from a simple estimate for the mean energy $E_J = \hbar I_c/2e$ using
the experimental value of the low-temperature, Z.F.C. ($T=10 K, H_{dc}=0$)
 critical current density  $J_c \approx 3.7 A/cm^2$.  Using the estimate
 $a_0 \approx 5 \mu m$ for the mean size of the grains, one  could naively
 obtain $I_c \approx
 J_c a_0^2 \approx 1 \mu A$ and the corresponding  low-temperature
 Josephson energy  $E_J^{naive}
 \approx 20 K $  (this value was derived from $J_c$ measured at $T=10 K$, but
 we do not expect much difference in the intrinsic Josephson energies at
 $T = 10 K$ and at $T \rightarrow 0$ since the bulk transition temperature
 in La$_{1.8}$Sr$_{0.2}$CuO$_4$~ is $T_c \approx 32 K$).
  However such an estimate is in contradiction with the measured
 value of the glass transition temperature $T_g \approx 29 K$.
   Indeed, let us assume that the  mean
 coordination number (number of ``interacting neigbours'') $Z$ in the ceramics
 is around 6, as for a simple cubic lattice.  Then for the
 estimate of the relation between $E_J$ and $T_g$ one can use the
 simulation data  \cite{young,huse} which give  $T_g \approx 0.5 E_J (T_g) =
 0.5 E_{J_o}\cdot (1-T_g/T_c)$, where we took into account the 
 linear dependence of $E_J$
 on $T_c-T$ close to the bulk transition temperature. As a result, one gets
 \begin{equation}
 \frac{E_{J_o}}{k_B} \approx \frac{2T_g}{1-T_g/T_c} \approx 600 K
 \label{E_J}
 \end{equation}
 i.e. a factor 30 larger than the naive estimate above. However we will show now
 that this discrepancy may be resolved if we assume that the current network
 producing the measured critical current density $J_c$ was actually
 {\it strongly frustrated} in spite of the absence of background d.c. field
 in this measurement.

 The  macroscopic critical current density $J_c$ for a strongly frustrated
 Josephson network was calculated in Ref.~\cite{Feigel} within the mean-field
 approach (we are not aware of any calculations of this kind beyond
 the scope of the mean-field theory). It was shown that frustration
 strongly reduces $J_c$
 as compared to its value $J_0$ for an unfrustrated system,
 $J_c/J_0 = \frac{3\sqrt{3} \gamma}{8} (1-T/T_g)^{5/2}$, where the factor
  $\gamma \approx 0.065$ was obtained by numerical solution of the
  {\it slow cooling} equations \cite{Vinok,Lev,Dotsen} describing
  the evolution of the glassy state under slow variations of
  temperature and magnetic
  field. In the low-temperature limit, this relation amounts to a factor 25
  reduction of the $J_c$ value with respect to $J_0$. Correspondingly, the
  characteristic value
  of the  critical current for an individual junction will be obtained as
  $I_c \approx 25 J_c a_0^2 \approx 25 \mu A$ and  results in a Josephson
  coupling energy  $E_{J_o} \approx 500 K$, in a fairly good agreement with the
  above estimate (\ref{E_J}).

  The above estimates show that the network of Josephson junctions in La$_{1.8}$Sr$_{0.2}$CuO$_4$~
  is frustrated even in the absence of an external magnetic field.
  A careful reader could  question this conclusion since we have
  used some results from
  the mean-field theory which may be a poor approximation for a
  3D gauge-glass.
  We believe, however that the qualitative result of the above estimates is
  sufficiently robust because a strong reduction of
  $J_c$ with respect to $J_0$ should  be a general feature of a glassy
  network, so that unaccuracy due to mean-field approximation
 cannot compensate for
  a huge discrepancy obtained between $E_J^{naive}$ and the estimate
(\ref{E_J}).
  Additional evidence in favor of the  glassy nature  of our system
  is provided  by the
  similarity of the low-$\Delta H$ diamagnetic response
  at $T=20 K$ with zero as well as non-zero $H_{dc}$ , as  described in Section III
  above, as well as the low-frequency noise data obtained in 
Ref.~\cite{Leylek2}
  on the same type of ceramics.

 What could be the origin of that frustration?
  We believe that most probably it is the result of the d-wave nature of
  superconductivity in
  cuprates \cite{d-wave} and randomness of the crystalline orientations
  in ceramics  \cite{gesh,sigrist}. It was shown there that
 the form of the effective phase-dependent Hamiltonian for such 
 ceramics  is of the same form as in (\ref{ham}) except for the fact
 that the random phases $\alpha_{ij}$ at $B=0$  are just $0$ or $\pi$ depending on the mutual orientation of grains $i$ and $j$.  Therefore such a system at $B=0$
 is equivalent to the $XY$ spin glass, with the low-temperature state
characterized by a completely random orientation of phases $\phi_i$, as in
the gauge-glass model with uniformly random distribution of $\alpha_{ij}$'s. 

Therefore the low-temperature state is characterized by
 the presence of randomly distributed intergrain currents
and, therefore, of the magnetic field generated by these currents. It means that
 the actual phases $\alpha_{ij}$ will contain contributions due to the self-induced magnetic field. Its relative importance
is characterized by the ratio of the corresponding magnetic flux penetrating elementary loops of the ceramics $\Phi_{sf}$ to the flux quantum $\Phi_0$, i.e.
just by the parameter $\beta_L =2\pi {\cal L }I_c/c\Phi_0$ where ${\cal L}$
is the characteristic inductance of an elementary loop \cite{Vela}.
Estimating the elementary inductance as ${\cal L} \approx 2\pi a_0 \mu_g$
and using Eq.(\ref{E_J}) to estimate $I_c$, we obtain
\begin{equation}
\beta_L \approx \frac{4\pi^2 a_0 \mu_gI_c}{c\Phi_0} =
 \frac{8\pi^3 \mu_g a_0 E_J}{\Phi_0^2} \approx 0.1
\label{beta}
\end{equation}
so the self-field (screening) effects are relatively weak, though perhaps not 
always negligible.

It is also of interest to estimate the effective penetration depth
$\lambda_{cer}$
 of a very weak magnetic field perturbation $\delta H$ into the ceramics.
Roughly, the value of $\lambda_{cer}$ can be estimated as
$a_0/\sqrt{\beta_L} \sim 15 \mu m$.  Another (hopefully more accurate)
estimate can be
obtained using mean-field results \cite{Feigel} which allow one to
express $\lambda_{cer}$ via the critical current density $J_c$:
\begin{equation}
\lambda_{cer} =
\left(\frac{\gamma}{8\pi^2}\frac{c\Phi_0}{J_c\xi_0\mu_g} \right)^{1/2} \approx
 25 \mu m
\label{lambda}
\end{equation}
where we inserted (as compared with Ref.~\cite{Feigel}) $\mu_g \approx 0.35$
and approximated the random nearest-neigbour network by a cubic lattice with
coordination number $Z=6$, which amounts to the relation
$\xi_0^2 = a_0^2/6 $ between the effective interaction range $\xi_0$ and
the intergrain distance $a_0$.

The characteristic magnetic field variation producing the critical current
density $J_c$ at the boundary can be estimated as
$\Delta H_c \sim 4\pi\lambda_{cer}J_c/c \approx 15 mG$,
whereas the numerical solution
\cite{Feigel} gives
\begin{equation}
\Delta H_c = \frac{\upsilon}{2c\gamma}4\pi\lambda_{cer}J_c \approx 30 mG
\label{H_c}
\end{equation}
Within the theoretical approach of Ref.~\cite{Feigel},  $\Delta H_c$ 
marks a crossover
between reversible (although  still non-linear at $\Delta H \leq \Delta H_c$)
and irreversible penetration of the magnetic field into the intergrain network.
The value of $\Delta H_c$ obtained in Eq.(\ref{H_c}) is on the lower border of the range of the field variations used to measure our d.c. magnetization 
curves, so we could just conclude
that we always have $\delta H \gg \Delta H_c$ and thus are producing
the Bean-like critical state.
Indeed, the data at $H_{dc}=0$, $T=10K$ look compatible with such an interpretation
(cf. Fig.(\ref{dx2_1020}), where some deviations from the logarithmic slope 1
(which is the characteristic of a Bean state) are seen at lowest
$\Delta H \leq 50 mG$).

 However, as far as the data obtained at $10\,K$ with d.c. fields 
$H_{dc} \simeq 2G\,$, or all data at higher temperature ($T=20 K$), 
including d.c. and a.c. results at zero-$H_{dc}$, are concerned (cf. 
 Figs.(\ref{xinis_10}-\ref{xpjvsla})),
 the low-field magnetization response is drastically different
 from Bean-type predictions, as explained at the end of Section III.
Qualitatively, the most surprising feature of these data is the existence
of a very broad range of $\Delta H$ within which the response is non-linear
but still not like the
critical-state one.
 We are not aware of any microscopic theory which predicts
 fractional-power
 behaviour of the shielding susceptibility over such a broad range of $\Delta  H$
 variations.   It cannot be excluded {\it a priori} that such a behaviour is related
  to a very wide range of intergrain critical currents, which might exist in ceramics (till now we have neglected
  inhomogenity of intergrain
  coupling strengths in our theoretical discussion). Moreover, we may expect 
 that the relative importance of such inhomogenities increases with the field 
 and/or temperature (cf. Ref.~\cite{lebed})

In  Section V,
  we will try to formulate a  new phenomenological model appropriate for the
  understanding of our data (leaving its theoretical justification for a future
  study); this model will be seen to be an
  interpolation between Campbell's and Bean's regimes of flux penetration into
  hard superconductors.

\section{Fractal model of diamagnetic response}

We showed at the end of Section III-C that a simple generalization,
Eq.(\ref{dBdx2}),
of Bean's relation between variation of the applied magnetic induction
$\Delta B$ and current $J$ results in reasonably good agreement
with our data. However, contrary to the original Bean relation, the new
one was not based on any physical picture; it was just a
convenient description of the data. In this Section we propose a 
 phenomenological model which provides a qualitative
understanding of the irreversible diamagnetic behaviour mimicked by
Eq.(\ref{dBdx2}).

We start from the picture of  non-linear  response of the current ${\bf J}$
to a variation of the vector potential $\delta {\bf A}$
 derived in Ref.~\cite{Feigel} within the mean-field approximation and
presented in Fig. 2 of that paper. Here the current induced by a variation
of $\delta {\bf A}$ is linear at very small $\delta {\bf A}$, then grows sublinearly,
and finally reaches its maximum value $J_c$ at the critical $\delta A_c$
 such that the differential response
$(dJ/dA)_{\delta A_c} \rightarrow 0$.  At $\delta A > \delta A_c$
the numerical instability of the {\it slow cooling} equations was detected
and interpreted as an indication of the absence of any solution which would
interpolate smoothly between zero and large (i.e. $ \gg \delta A_c$)
values of $\delta A$. In other terms, some kind of ``phase slip'' was expected
to happen in the model \cite{Feigel}, leading to a new metastable state, which
would have lower (free) energy at the new value of the vector potential
${\bf A}' = {\bf A}+\delta {\bf A}$ (in other terms, a state similar to the one
obtained by the F.C. procedure at constant ${\bf A}'$, which
does not carry macroscopic current).
 Further increase of $\delta {\bf A}' = {\bf A} - {\bf A}'$ again
 induces a macroscopic current until it reaches the maximum value $J_c$
 at $\delta  A' = \delta A_c$, and so on. Thus the whole $J(\delta A)$
 dependence emerging from the mean-field solution \cite{Feigel} is periodic;
it leads immediately to the irreversibility of the response, since the inverse
function $\delta A (J)$ is multivalued: different vector potential values
may correspond to the same value of current. Of course, such a periodic 
$J(\delta A)$ dependence does not correspond to the usual Campbell - Bean picture, 
which would better be represented by 
\begin{equation}
J_{C.-B.}(\delta A)= J(\delta A)\theta (\delta A_c -\delta A ) 
+J_c\theta (\delta A -\delta A_c)
\label{Jbean}
\end{equation} 
It is important to note that the $J(\delta A)$
dependence Ref.~\cite{Feigel} was obtained from the {\it space-independent}
solution for the glassy correlation function
$Q_{jj}(t,t') = <cos(\phi_j(t)-\phi_j(t'))>$; such an approximation,
being  reasonable for the description of smooth ``adiabatic'' transformations
in a system with long-range interactions, will probably break down when
the jump from one metastable state to another happens. In other terms,
the above-mentioned ``phase slip'' should have something to do with spatially
inhomogeneous processes like vortex penetration in hard type-II
superconductors.  The problem of the solution of the general history- and
space-dependent system of 
integral equations (which may be derived following the method
 of Ref.~\cite{Feigel})
is formidable and the method to solve it is still unknown. Therefore we
can only speculate on possible properties of its solution. The simplest idea
would be that the macroscopic $J(\delta A)$ response becomes (after averaging
over inhomogenities of the space-dependent solution) similar to the Campbell-Bean type of the
response (\ref{Jbean}).  Indeed, our analysis of the low-field diamagnetic
response at $T=10 K$ and $H_{ext}=0$ (Section III-B) developed in Section IV
on the basis of such an assumption, is in reasonable agreement with the data.
However other sets of data (for higher temperature and/or lower field) are
 described by completely different {\it Ansatz} (\ref{JBnew}).
We will now propose a (phenomenological) generalization of the $J(\delta A)$
relation compatible with Eq.(\ref{JBnew}).  The relation
we are looking for should be an {\it intrinsic} 
(i.e. independent on the sample geometry) and {\it general} 
(i.e. usable for an arbitrary magnetic history of the sample) relation 
between the current and variation of the vector potential. Remember that 
Eq.~(\ref{JBnew}) was written for the 
simplest nonmonotonic variation of $\Delta B$, and that it relates 
the true vector
$\bf J$ and the pseudovector $\Delta\bf B$. So, in writing this equation,
  some additional information on the geometry of the sample has been used 
  (we use the simplest slab geometry). Thus a natural basic equation should 
  relate the current density $\bf J$ and the variation of the vector
potential $\delta \bf A$.

In a generalized model, the diamagnetic current response should possess two major properties: \\
i) it must scale as some fractional power $\alpha
\approx 0.5$ with the amplitude of exitation field $\delta B$, and \\
ii) it must be strongly irreversible (as it follows from the analysis of the ratio
$4\pi\chi''/(1-4\pi\chi') \approx 0.28$ ). We consider these two conditions in
sequence.

The condition i) is rather easy to fulfill: it is enough to
suppose that the differential response of the current to the variation of the
vector potential $\delta A$ is given by a non-linear generalization of the
London relation
\begin{equation}
\frac{d{\bf  J}}{d{\bf A}} = -\frac{c}{4\pi}{\lambda_{eff}^{-2}(J)}
\label{londgen}
\end{equation}
where the current-dependent ``effective penetration depth'' is given by
\begin{equation}
\lambda_{eff}=\lambda_1 {\vert J/J_1 \vert}^{\kappa}\,.
\label{lambdaef}
\end{equation}
In the case of a monotonic field variation applied to an initially uniform
induction distribution, the Eqs.(\ref{londgen},~\ref{lambdaef}) lead
to the simple
relation $J \propto {\Delta B}^{\alpha}$ with $\alpha = (1+\kappa)^{-1}$.
Indeed, with $dA = \Delta B\,dx$ and approximating $d\Delta B/dx$ by
$\Delta B/\lambda_{eff}\,$, one obtains
$ J \propto {\Delta B}^{1\over {1+\kappa}}\,$.
Thus we need to choose $\kappa \approx 1$ in order to reproduce the observed
scaling with $\alpha \approx 0.5$.

However, the set of equations (\ref{londgen},~\ref{lambdaef}) does not fulfill
the second condition ii) above: the corresponding solutions are
{\it reversible}, as it follows from the existence of a single-valued
function $\delta A(J) \propto J^{1+2\kappa}$ which follows from
Eqs.(\ref{londgen},~\ref{lambdaef}). In other words, the system described by
Eqs.(\ref{londgen},~\ref{lambdaef}) would exhibit nonlinearity and harmonics
generation, but would not show finite $\chi''(\omega)$ in the $\omega
\rightarrow 0$ limit. In order to avoid this inconsistency, we need
to formulate a model with the same kind of scaling between $\delta A$ and
$J$ as in Eqs.(\ref{londgen},~\ref{lambdaef}), but with a nonmonotonic 
$J(\delta A)$ dependence allowing for the irreversible behaviour.

A model obeying
very similar properties  was formulated and studied in Ref.~\cite{1dSG} 
in a different physical context (one-dimensional spin glass). The low-energy  spin configurations in this model are
described by a phase variable $\varphi (x) \in (-\pi,\pi)$ such that
two such configurations (local energy minima) which differ by a phase shift
 $\delta \varphi (x_0) = \Phi$ in a region around some point $x_0$,
have a characteristic energy difference $E(\Phi) \propto \Phi^{5/3}$ and
a characteristic spatial extent of the phase deformation
$X(\Phi) \propto \Phi^{1/3}$. This scaling holds for the intermediate
range of phase deformations $\varphi_0 \ll \Phi \ll \pi$; at smaller
$\Phi \leq \varphi_0 \ll 1$ the energy cost of deformation is $\propto \Phi^2$,
whereas at $\Phi \sim \pi$ the energy growth obviously saturates due to
$2\pi$ periodicity. The above $E(\Phi)$ scaling leads to a {\it sublinear
growth} of the characteristic ``force'' $f(\Phi) = dE/d\Phi \propto \Phi^{2/3}$
with $\Phi$ in the same intermediate range.
The main contribution to the
second derivative $d^2E/d\Phi^2$ (curvature of the energy valleys) comes from
the smallest scale $\Phi \sim \varphi_0$, i.e. from the curvature of individual
local minima. It was explained
in Ref.~\cite{1dSG} that such a scaling means a fractal organization of the energy
minima as a function of $\varphi$ with fractal dimensionality $D_f = 1/3$.
It means that the number of energy minima discernable on a scale $\varphi$
grows as ${\cal N} \propto \varphi^{-1/3}$ at finer scales;
 new minima appear primarily due to the splitting the older (broader)
ones. This picture emerged in Ref.~\cite{1dSG} from the microscopic
analysis of the original Hamiltonian for a one-dimensional spin-glass model
formulated in Ref.~\cite{1d1}. We can borrow the qualitative features of
this construction for our present purpose (leaving for future studies
the problem of its microscopic justification for the case of superconductive
glasses).

Suppose that the free energy $F(\delta A)$
of the Josephson network behaves (as a function of  vector potential variations
with respect to a ``virgin'' state with a homogeneous induction)
in a way similar to  $E(\varphi)$ at $\varphi \ll \pi$. Namely, suppose that
the free energy is parabolic,
$\delta F \propto (\delta A)^2$, at very small
variations of vector potential $\delta A \leq \delta A_{c1}$, but on a larger scale,
$\delta A \gg \delta A_{c1}$, it
 contains many local minima whose characteristic free energies scale
(with respect to the lowest state with $\delta A =0$) as
\begin{equation}
F(\delta A) \propto (\delta A)^{\theta + 1}\, for  \delta A_{c1} \ll \delta A \ll\delta A_{c}
\label{F_A}
\end{equation}
with the exponent $\theta \in (0,1)$ (see the definition of $\delta A_{c}$
below). Then the characteristic value of the
current $J = \frac{1}{c}\partial F/\partial A$ scales as
\begin{equation}
J_{char}(\delta A) \approx J_{c1} \left(\frac{\delta A}{\delta A_{c1}}\right)^{\theta}
\label{J_A}
\end{equation}
in the same interval of $\delta A$. At large $\delta A \geq \delta A_c$
variations, the growth of the induced current should saturate at the true
critical current value $J_c$, so we can estimate
\begin{equation}
\delta A_c \sim \delta A_{c1} (J_c/J_{c1})^{1/\theta}.
\label{A_c}
\end{equation}
On the other hand, weak
$\delta A \ll \delta A_{c1}$ leads to the usual linear London (or Campbell)
response with an effective penetration depth $\lambda_1$; matching at
$\delta A \sim \delta A_{c1}$ leads to the following estimate:
\begin{equation}
\delta A_{c1} \sim \frac{4\pi}{c}J_{c1} \lambda^2_1
\label{A_c1}
\end{equation}
The estimate (\ref{J_A}) looks very much like the previous version defined by
(\ref{londgen},~\ref{lambdaef}), so one can find the relation between the
exponents:
\begin{equation}
\theta = 1/(1+2\kappa) = \alpha/(2-\alpha) \approx 0.3
\label{theta}
\end{equation}
\begin{figure}[htb]
\epsfxsize=\hsize \epsfbox[0 0 400 300]{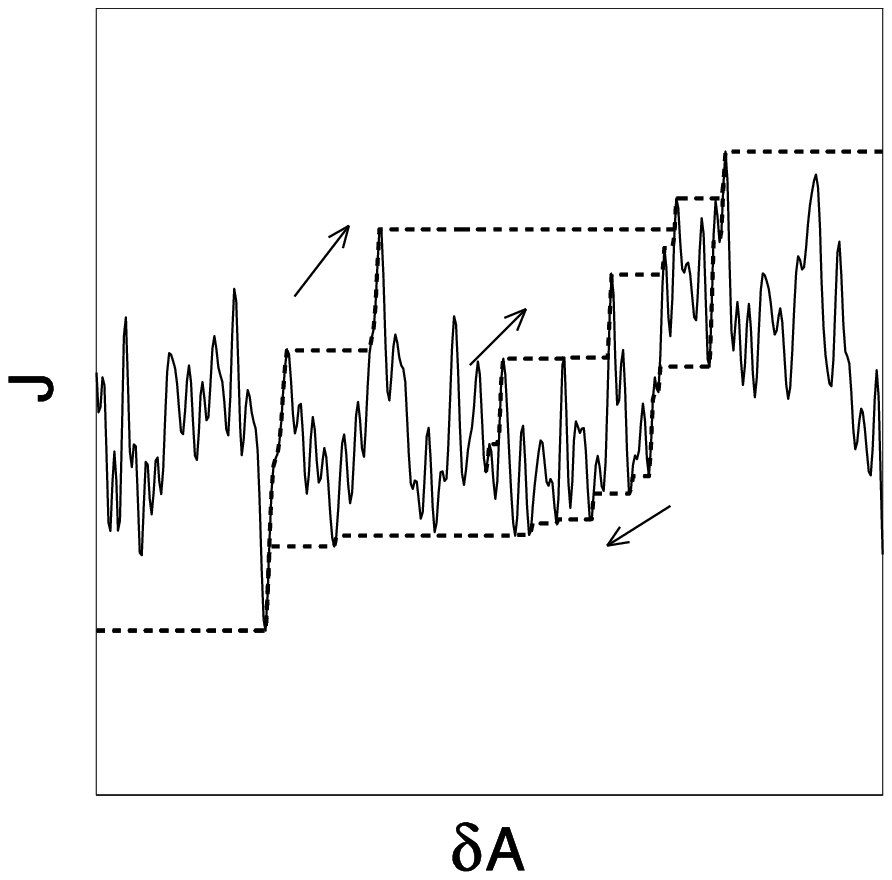}
\caption{Picture of a fractal $J(\delta A)$ landscape. An example of 
a hysteresis loop is shown.}
\label{simulj}
\end{figure}

However the whole picture is substantially altered: the current is now
supposed to be an (irregularly) oscillating function of $\delta A$ 
(see Fig.~\ref{simulj}),
thus only its envelope $J_{char}(\delta A)$ defined on a scale $\delta A$  follows
the scaling relation (\ref{J_A}). As a result, the inverse function
$\delta A(J)$ is multivalued and the irreversibility of the response is
 ensured. Similar to the spin-glass model of Ref.~\cite{1dSG}, the fractal dimensionality
$D_f$ of the low-energy valleys can be defined; it is given now by
$D_f = 1 - \theta \approx 0.7$.  The proposed picture is based on the existence of
two substantially different scales of currents, $J_{c1}$ and $J_c$, and
corresponding vector potential variations $\delta A_{c1}$ and $\delta A_c$;
thus it can be compared with the usual Campbell-Bean picture of
critical currents in the same way as the thermodynamics of
 type-II superconductors is compared with that of the type-I ones. 

In order to describe  quantitatively the diamagnetic response in the ``fractal''
range (\ref{F_A}) we need to determine the {\it distribution function}
${\cal P}[J(\delta A)]$ (which would lead, in particular, to the estimate
(\ref{J_A}) for $J_{char}(\delta A)$). Moreover, in general, a relation
of the type of (\ref{J_A}) could be nonlocal (i.e. the current depends on the
$\delta A(x)$ distribution in some region of space, whose size may depend
on $\delta A$ itself (see again Ref.~\cite{1dSG}). We leave this complicated problem
for future studies, and just note here that  merely the existence of
relation (\ref{J_A}) is sufficient for the existence of some ``natural'' properties of the response (like the presence of a closed hysteresis loop,
as it was assumed in Section III-C).

\section{Summary and conclusions} 
In this paper, we have presented experimental results on the low temperature 
($10\,K$ and $20\,K$) response of the granular HT$_c$ superconductor La$_{1.8}$Sr$_{0.2}$CuO$_4$ 
 to small field excitations. The general properties of the magnetic response were investigated 
in two samples (A and B) differing by the strength of the coupling between grains. 
By cooling the samples in various d.c. fields up to $20\,G$ and applying 
small field increases, we were able to measure the shielding response of 
the material and to derive a method, inspired by the work of Dersh and 
Blatter \cite{Dersh}, to extract from the data the 
polarizability of the intergrain currents system. The field cooled (F.C.) 
magnetization was measured in fields up to $20\,G\,$. Analysis of the results 
led to the conclusion that i) the structure of the grains is polycrystalline, 
resulting in a step decrease of the F.C. magnetization with increasing 
field, which can be interpreted on the basis of the model by Wohllebeen et al. \cite{Wohll}; 
ii) self shielding (pinning) by the intergrain currents when lowering the 
temperature strongly reduces the value of the F.C. magnetization; iii) there 
is no macroscopic Meissner magnetization due to the system of intergrain 
currents. 

Further detailed study of the response of the Josephson network was performed 
in sample B. It was shown that the response is asymmetric with respect to the sign  of  variation of the 
applied field after field cooling; this is due to the shielding currents pinned 
during cooling.  The macroscopic critical current is found to be
strongly reduced by moderate values of the external d.c. field, 
about $2\,G\,$. 

Very low field magnetization measurements were performed by applying field 
steps of $10\,mG$ or low frequency a.c. fields in the range $50\,\mu G$ to 
$30\,mG$, after cooling in d.c. fields up to $8.8\,G\,$. The results show that 
the response is strongly non linear, the shielding current growing 
{\it sublinearly} with increasing applied field. Furthermore, the a.c. 
results show that it is strongly {\it irreversible} down to the smallest 
excitations used. It is shown that a non-linear relation between the 
shielding current and the induction, $J\propto \Delta B^{\alpha}$ with 
$\alpha \approx 0.5$, together with a natural assumption about the existence
of a closed hysteresis loop,
 give predictions in a reasonable agreement 
with the data. 

Theoretical analysis of our experimental results was developed on the frame 
of the existing ``gauge-glass" theories. It was show that the extremely low 
value of the low-temperature, zero-field critical current density 
($J_c\approx 3.7\,A/cm^2$ at $10\,K$) together with the rather high temperature of the 
transition to the low-temperature glassy state, can be coherently interpreted 
only under the assumption that the Josephson network is strongly frustrated even
at zero applied field. This contradicts the usual assumption that frustration 
in the interactions arises only due to the local magnetic induction, 
but supports the hypothesis of the existence of a large proportiont of 
$\pi$-junctions in the granular system. These $\pi$-junctions are possibly due 
to the d-wave nature of the pairing, combined with the randomness
of grain orientations in La$_{1.8}$Sr$_{0.2}$CuO$_4$~ ceramics.

Finally, a new model of diamagnetic response in the glassy state of granular 
superconductors was developed in order to describe the anomalous (fractional-power) behavior 
of the shielding current response. This model, based on the idea of a fractal 
organization of the free energy landscape in the granular network, can provide  
a qualitative account for the main features of the anomalous response. 
Its further development will be the subject of future studies. 

\section{Acknowledgements}

We are grateful to L. B. Ioffe for many important discussions which helped to clarify a number of issues considered in this paper. 
Research of M.V.F. was
 supported by the DGA grant \# 94-1189,
  by the joint grant
M6M300 from the International Science Foundation and the Russian Goverment,
and by the grant \# 95-02-05720 from the Russian Foundation for Fundamental
Research.

\appendix
\section{Appendix}
\begin{figure}[htb]
\epsfxsize=\hsize \epsfbox[0 0 460 300]{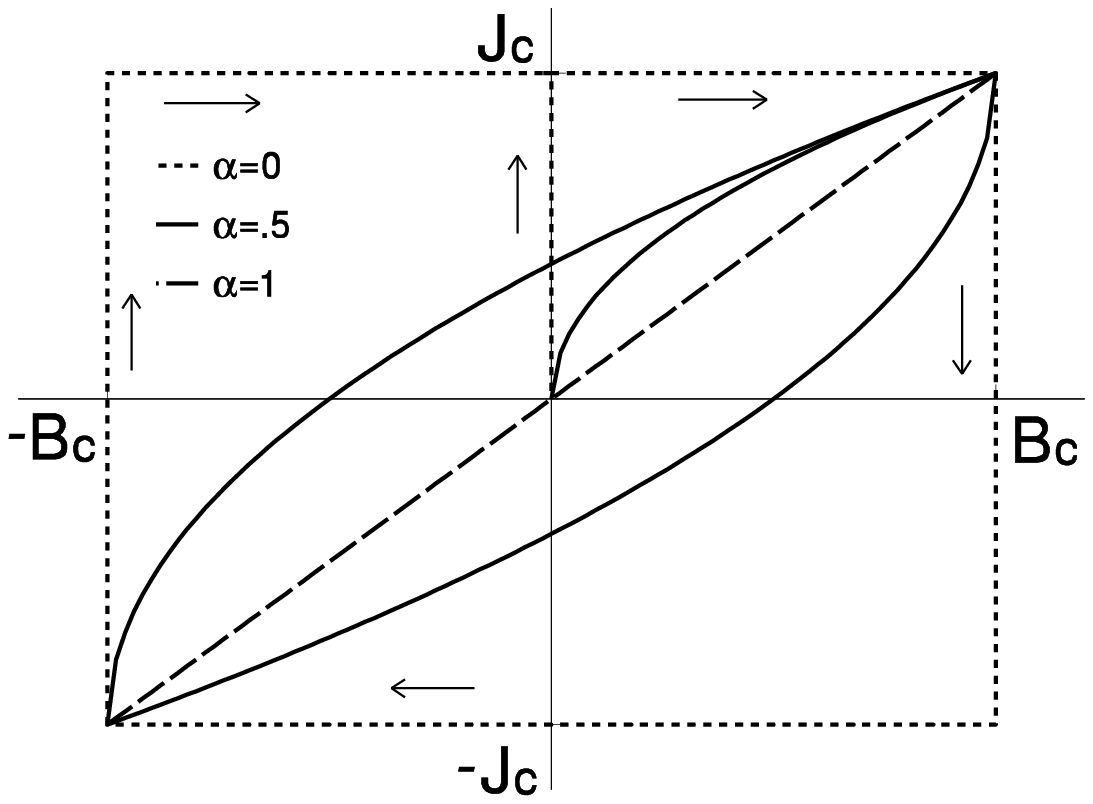} 		 
\caption{Hysteretic behavior of the current for a variation of the induction 
between $-B_c$ and $B_c\,$.}
\label{jdeb}
\end{figure}

The hysteretic behavior of the current as a function of the induction 
variations is represented by the relation:
\begin{equation}
\Delta J = \pm 2^{(1-\alpha)\nu} J_c \left({\vert\Delta B\vert}\over B_c\right)^{\alpha}
\label{djmod}
\end{equation}
$\nu =0$ when starting from zero induction state, and 1 otherwise. 
$\Delta J =J-J_o$ and $\Delta B=B-B_o$ where $J_o$ and $B_o$ are the (old) 
values just before the last reversal of the sign of variation of $B\,$. 
The {\it Ansatz}~\ref{djmod} ensures that we have a stable closed hysteresis 
loop, and that there is no hysteresis for $\alpha =1$ which describes the 
London case. Figure~\ref{jdeb} displays the shape of the hysteresis loops 
for three characteristic values of $\alpha\,$. 
The induction profile is determined by the Maxwell equation which leads,
for the case of weak penetration, to  
\begin{equation}
{{d\Delta B }\over dx}=\pm 2^{(1-\alpha)\nu} {\cal A}\,\vert{\Delta B}\vert ^{\alpha}
\label{dBdxa} 
\end{equation}
where ${\cal A}=4\pi J_c /B_c^{\alpha}\,$; $x$ is the distance from the edge of the sample. After increasing applied field from 0 to $h_o$, starting 
from zero induction state, the induction profile is given by 
$B^{-\alpha}dB = -{\cal A}\,dx\,$, leading to:
$$x=-{1\over {\cal A}}\,\int^B_{h_o}\,\xi^{-\alpha}d\xi= 
-{{B^{1-\alpha}-h_o^{1-\alpha}}\over {(1-\alpha){\cal A}}}$$ 
where
\begin{equation}
B=\left(h_o^{1-\alpha} - (1-\alpha){\cal A}\,x \right) ^{1\over{1-\alpha}}. 
\label{initho}
\end{equation} 
Field penetrates till $x=x_{h_o}=h_o^{1-\alpha}/(1-\alpha){\cal A}\,$. 

\begin{figure}
\epsfxsize=\hsize \epsfbox[0 0 400 300]{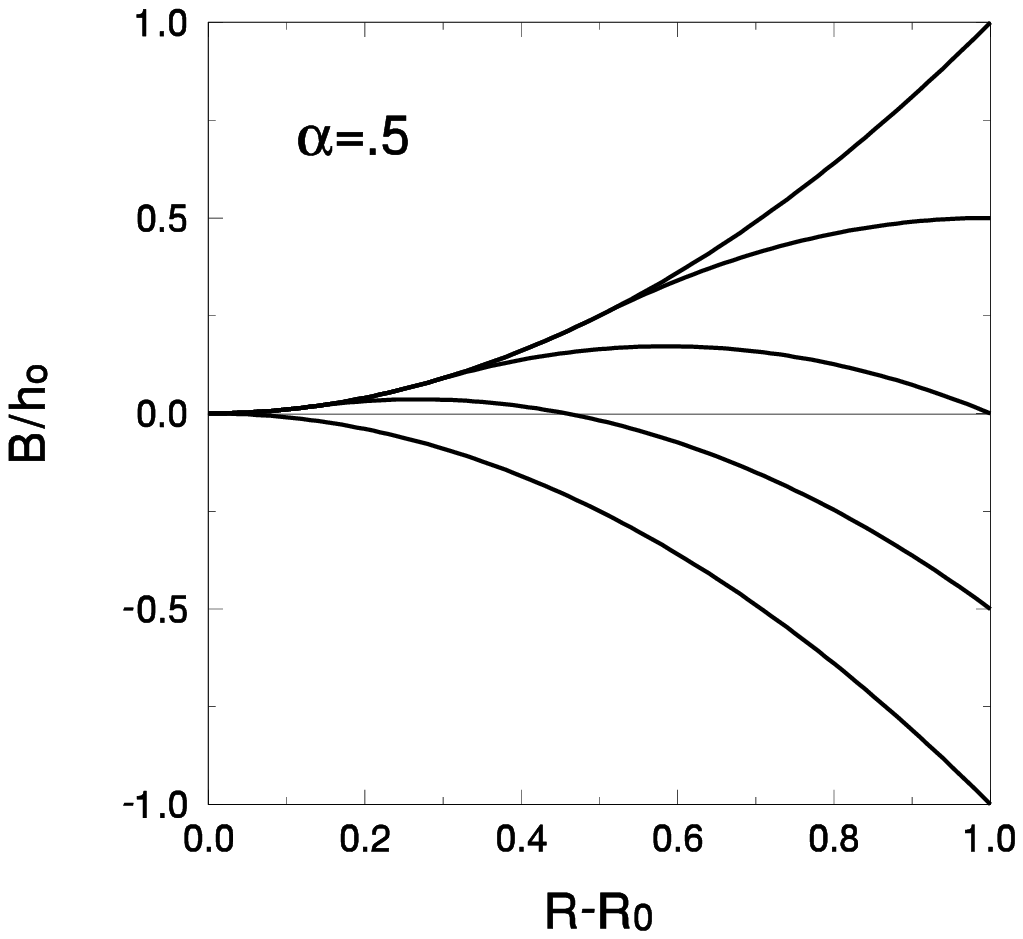}
\caption{Induction profile as a function of applied field. $R_o$ is the 
maximum abcissa of field penetration.} 
\label{bprofil} 
\end{figure}

When $h$ decreases from $h_o$, we get $(B_o -B)^{-\alpha}dB=-2^{1-\alpha}dx\,$. 
Hence:
\begin{eqnarray*}
x=-{1\over {2^{1-\alpha}{\cal A}}}\,\int^{B_o -B}_{h_o -h}\xi^{-\alpha}d\xi 
\nonumber\\ 
=-{1\over{2^{1-\alpha}{\cal A}(1-\alpha )}}\left( (B_o -B)^{1-\alpha}-(h_o -h)^{1-\alpha}\right) 
\end{eqnarray*} 
Modification of induction relative to $B_o$ extends up to 
$x_h = (h_o -h)^{1-\alpha}/2^{1-\alpha}{\cal A}\,$.
For $0<x<x_h\,$, 
\begin{equation}
B= B_o -2\left(\left({{h_o -h}\over 2}\right) ^{1-\alpha} - 
(1-\alpha){\cal A}\,x \right) ^{1\over{1-\alpha}} 
\label{reversh}
\end{equation}
where $B_o$ is given by Eq.~(\ref{initho}). When $h=-h_o$ is reached, 
Eq.~(\ref{reversh}) gives simply $B=-B_o\,$. After reversing the sign 
of variation of $h$ once more, the  
profiles are simply symmetrical of those given by Eq.~(\ref{reversh}).
Examples of induction profiles are given on Fig.~\ref{bprofil} for 
$\alpha =0.5\,$. 
The average induction can be derived now. After some algebra, one obtains:  
$$<B>={{\cal A}\over {2-\alpha}}\times h_o^{2-\alpha} \left[1- 
2\left({1-{h/h_o}\over 2}\right) ^{2-\alpha}\right]$$ 
for $h_o >0\,$, and 
\begin{equation}
<B>={{\cal }\over {2-\alpha}}\times h_o^{2-\alpha} \left[-1+ 
2\left({1+{h/h_o}\over 2}\right) ^{2-\alpha}\right] 
\end{equation}
for $h_o <0\,.$

\vspace{.2cm} 
For a sinusoidal excitation $h=h_o\,\cos\omega t$, one gets 
$${<B>\over {h_o}}={{{\cal A}\,h_o^{1-\alpha}}\over {2-\alpha}}\left[1-
2\left({{1-\cos\omega t}\over 2}\right) ^{2-\alpha} \right]$$
for $2n\pi<\omega t<(2n+1)\pi$
\begin{equation}
{<B>\over {h_o}}={{{\cal A}\,h_o^{1-\alpha}}\over {2-\alpha}}\left[-1+
2\left({{1+\cos\omega t}\over 2}\right) ^{2-\alpha} \right]
\end{equation}
for $(2n-1)\pi<\omega t<2n\pi\,.$

\vspace{.2cm}

\begin{figure}
\epsfxsize=\hsize \epsfbox[0 0 400 300]{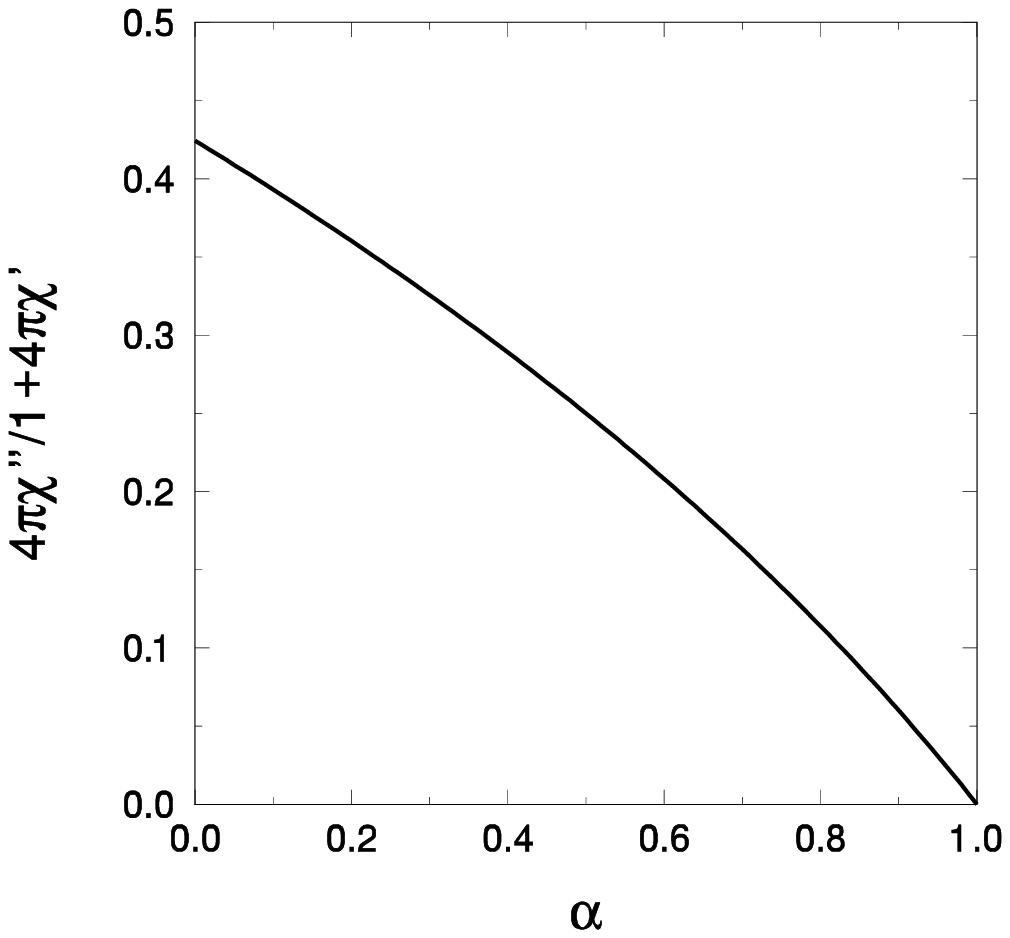}
\caption{Values of $4\pi\chi''/1+4\pi\chi'$ as a function of $\alpha\,$.}
\label{pentxsxp}
\end{figure}

Since ${<B>/{h_o}}=1+<M>/h_o\,$, Fourier transformation gives the values of
$1+4\pi\chi'$ and $4\pi\chi''\,$. This can be done numerically. 
Figure~\ref{pentxsxp} displays the ratio $4\pi\chi''/1+4\pi\chi'$ as a 
function of $\alpha\,$.


\begin{thebibliography}{99}
\bibitem{note}  The susceptibility, which in the
      usual sense is a local quantity representing $(B - H)/H\,$, has no meaning
      in the case of circulating currents in a conductor. We speak rather of a
      polarizability $\chi_j$ which represents
      the average value $<B-H>/H\,$, and describes the global effect of the
      currents over the whole volume of the sample.
\end{thebibliography}
\end{document}